\documentclass[fleqn,usenatbib]{mnras}

\usepackage{newtxtext}
\usepackage{mathptmx}

\usepackage[T1]{fontenc}

\DeclareRobustCommand{\VAN}[3]{#2}
\let\VANthebibliography\thebibliography
\def\thebibliography{\DeclareRobustCommand{\VAN}[3]{##3}\VANthebibliography}

\DeclareMathAlphabet{\mathcal}{OMS}{cmsy}{m}{n}

\usepackage{graphicx}	
\usepackage{amsmath}	
\usepackage{amssymb}	

\usepackage{lipsum}

\newcommand{\new}[1]{{#1}}

\newcommand\iona[2]{#1\,{\sc #2}}

\title[Gas flows in galaxy mergers]{Gas flows in galaxy mergers: supersonic turbulence in bridges, accretion from the circumgalactic medium, and metallicity dilution}

\author[Sparre et al. ]{Martin Sparre$^{1,2}$\thanks{E-mail: sparre@uni-potsdam.de}, Joseph Whittingham$^{2}$, Mitali Damle$^1$, Maan H. Hani$^{3}$\thanks{Herschel fellow}, Philipp Richter$^1$, \newauthor Sara L. Ellison$^4$, Christoph Pfrommer$^2$,  Mark Vogelsberger$^5$
\\
$^{1}$Institut f\"ur Physik und Astronomie, Universit\"at Potsdam, Karl-Liebknecht-Str.\,24/25, 14476 Golm, Germany\\
$^{2}$Leibniz-Institut f\"ur Astrophysik Potsdam (AIP), An der Sternwarte 16, 14482 Potsdam, Germany\\
$^3$Department of Physics and Astronomy, McMaster University, Hamilton Ontario L8S 4M1, Canada\\
$^{4}$Departmentof Physics and Astronomy, University of Victoria, Victoria, British Columbia, V8P 1A1, Canada\\
$^{5}$Department of Physics, Kavli Institute for Astrophysics and Space Research, Massachusetts Institute of Technology, Cambridge, MA 02139, USA
}

\pubyear{2021}

\begin{document}
\label{firstpage}
\pagerange{\pageref{firstpage}--\pageref{lastpage}}
\maketitle

\begin{abstract}
In major galaxy mergers, the orbits of stars are violently perturbed, and gas is torqued to the centre, diluting the gas metallicity and igniting a starburst. In this paper, we study the gas dynamics in and around merging galaxies using a series of cosmological magneto-hydrodynamical (MHD) zoom-in simulations. We find that the gas bridge connecting the merging galaxies pre-coalescence is dominated by turbulent pressure, with turbulent Mach numbers peaking at values of \new{1.6--3.3}. This implies that bridges are dominated by supersonic turbulence, and are thus ideal candidates for studying the impact of extreme environments on star formation. We also find that gas accreted from the circumgalactic medium (CGM) during the merger significantly contributes (27--51 per cent) to the star formation rate (SFR) at the time of coalescence and drives the subsequent reignition of star formation in the merger remnant. Indeed, 19--53 per cent of the SFR at $z=0$ originates from gas belonging to the CGM prior the merger. Finally, we investigate the origin of the metallicity-diluted gas at the centre of merging galaxies. We show that this gas is rapidly accreted onto the galactic centre with a time-scale much shorter than that of normal star-forming galaxies. This explains why coalescing galaxies are not well-captured by the \emph{fundamental metallicity relation}.
\end{abstract}



\section{Introduction}\label{Intro}

Galaxy mergers play an important role in galaxy formation. A major merger of gas-rich galaxies can, for example, cause a starburst \citep{1996ARA&A..34..749S,2013MNRAS.428.2529H}, where the star formation rate (SFR) is significantly enhanced in comparison to normal star-forming galaxies. Furthermore, minor and major mergers of gas-poor galaxies are understood to drive the observed size evolution of quenched elliptical galaxies \citep{2007A&A...476.1179B,2009ApJ...699L.178N,2012ApJ...744...63O,2017ApJ...834...18B,2020ApJ...888....4S}. Cosmological galaxy formation simulations also favour a large fraction of mass accreted through mergers for massive galaxies \citep{2013MNRAS.433.3297D,2016MNRAS.458.2371R}.

Idealised simulations of the collision of galaxies have given us a remarkable understanding of mergers. The presence of a strong merger-induced starburst is well-established \citep{1996ApJ...464..641M,2005MNRAS.361..776S,2008MNRAS.391.1137L, 2014MNRAS.442.1992H,2015MNRAS.448.1107M,2019MNRAS.490.2139R,2020MNRAS.493.3716H}, and the transformation of star-forming disc galaxies into quenched ellipticals has also been studied \citep{2008ApJS..175..356H}. Whether the fate of a disc galaxy merger is a quenched elliptical does, however, depend on the role of black hole feedback \citep{2005Natur.433..604D}, and the orbital parameters of the merging galaxies \citep{2005ApJ...622L...9S}.

The stars in merging galaxies experience \new{forces}, which perturb their orbits. This \new{frequently results in the formation of a bridge} connecting the galaxies, as well as tails of stars stripped from the disc. The generation of such phenomena has been established in simple test-particle simulations \citep{1963ZA.....58...12P,1972ApJ...178..623T,2009AJ....137.3071B,2013ApJ...771..120P}, and in more sophisticated simulations with self-gravitating stellar discs \citep[e.g.,][]{1992ApJ...393..484B}.  Visual inspection of the gas distribution in closed-box simulations of mergers also reveals bridges and tails \citep[see e.g. images in][]{2006MNRAS.373.1013C,2015MNRAS.446.2038R,2019MNRAS.485.1320M}. Such closed-box simulations do, however, initialise all the gas in the simulation to be inside the galaxies' discs, \new{meaning} they do not model the contribution of gas from the circumgalactic medium (CGM), which is the region outside the galaxy's disc, but still within \new{its dark matter halo}.

Our understanding of gas flows in and around galaxies has substantially improved in the last decade. Various observational campaigns have been initiated to map the gas in the CGM, showing it to be a critical component of the galactic ecosystem. Absorption studies with the \textsc{HST/COS} instrument have revealed the multiphase structure of the CGM gas, which makes up a significant fraction of a halo's baryon budget \citep{2014ApJ...792....8W,2017ARA&A..55..389T,2017A&A...607A..48R}. Furthermore, simulations reveal a rich structure in the CGM \citep{2019ApJ...873..129P,2019MNRAS.482L..85V,2019ApJ...882..156H,2019MNRAS.488..135H,2020MNRAS.498.2391N}, as well as a continuous gas transfer between the interstellar medium (ISM), the CGM, and the cosmological surroundings \citep{2017MNRAS.470.4698A,2019MNRAS.483.4040S,2019MNRAS.488.1248H, 2020MNRAS.494.3581H}. Idealised merger simulations, therefore, miss an important component of the picture of galaxy interactions.

Studies of mergers in cosmological simulations hint that gas transfer to the ISM during a merger plays an important role. For example, the merger remnants in \citet{2017MNRAS.470.3946S} are star-forming and a fraction of them have a blue disc. This is only possible because of the accretion of gas after the merger. This is supported by work by \citet{2018MNRAS.475.1160H}, who found signatures in a separate cosmological merger simulation that imply a mix of inflows and outflows in the CGM. This trend is also seen in observational work, such as that by \cite{2014MNRAS.445.2198O}, who find enhanced gas accretion in high-redshift mergers.

The aim of this paper is to quantify the gas flows in cosmological merger simulations. We study the gaseous bridge and determine the contribution of turbulent, thermal and magnetic pressure components, and we investigate whether \new{the gas in the bridge} originates from the CGM or whether it is purely populated by stripped disc material. We also track the gas residing in the CGM before the merger, and determine the fraction of merger-induced star formation whose gas supply originates from the CGM, rather than from the original galactic disc. Finally, we track gas which gives rise to an epoch of \emph{metallicity dilution} in the mergers. We present the simulations and methods in Sec.~\ref{Methods}, our results are presented in Sec.~\ref{Results}, and the discussion and conclusion in Sec.~\ref{Discussion} and Sec.~\ref{Conclusion}, respectively.

\begin{figure*}
\centering
\includegraphics[width=0.8\linewidth]{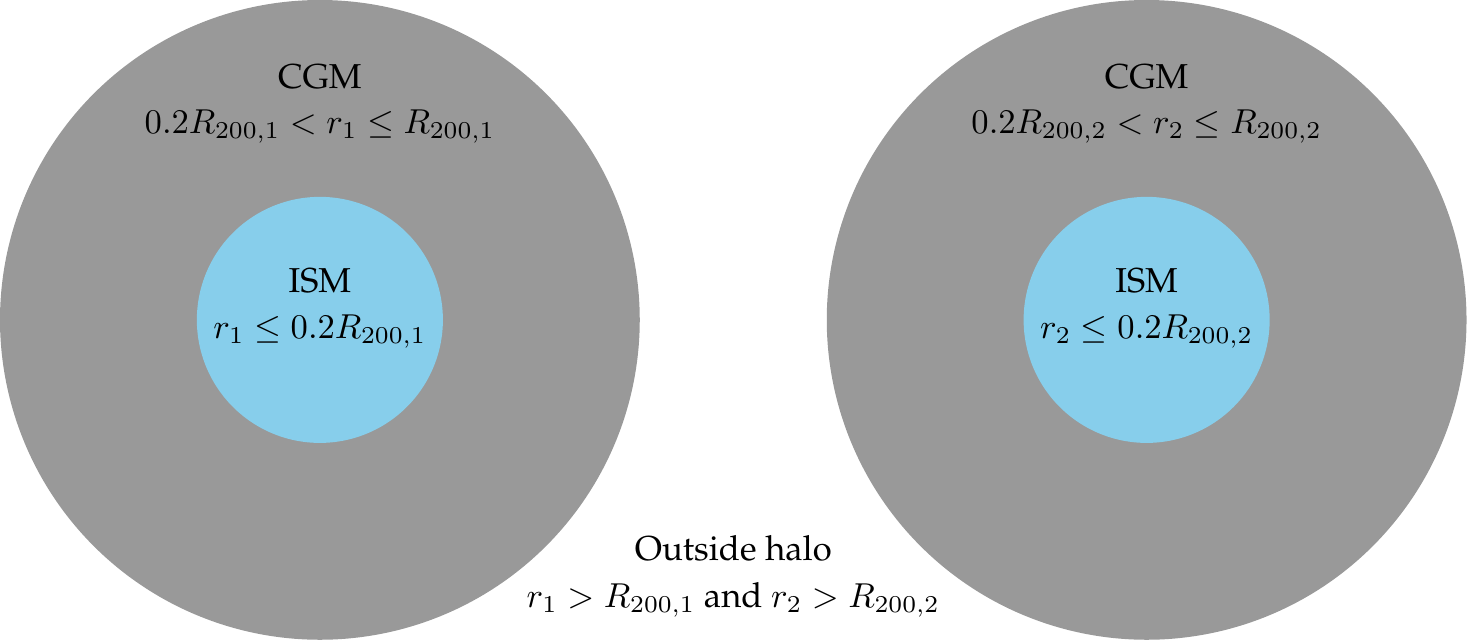}
\caption{Each gas cell is tagged as belonging to one of the three gas reservoirs: the ISM (if a gas cell is within $r\leq 0.2R_{200}$ of one of the merging galaxies), the CGM ($0.2 R_{200}<r\leq R_{200}$), or as not belonging to any of the merging galaxy haloes ($r>R_{200}$). We use these definitions to trace the origin of the gas mass budget in bridges in Fig.~\ref{Fig100_HI_sequence_TracedGasInBridge}, and for determining the origin of the star-forming gas in Fig.~\ref{MS_PlotTracedSFR}.}
\label{ISM_CGM_Definition}
\end{figure*}

\section{Methods}\label{Methods}

\subsection{Simulations}\label{SubSecSimulations}

In this paper, we analyse the four cosmological zoom simulations from \citet{2020arXiv201113947W}. These simulations are run with the code \textsc{arepo} \citep{2010MNRAS.401..791S,2016MNRAS.455.1134P,2020ApJS..248...32W}, which models magnetohydrodynamics (MHD) on a moving Voronoi mesh. The simulations have a dark matter mass resolution of $1.64 \times 10^5$ M$_\odot$, and a dark matter softening of 0.22 kpc. The mass of gas cells is kept within a factor of two of the baryonic target mass, which is $2.74 \times 10^4$ M$_\odot$. The initial conditions of the simulations were selected from the Illustris simulation \citep{2014MNRAS.444.1518V,2014Natur.509..177V,2014MNRAS.445..175G}, such that each galaxy had a major merger (with a merger mass ratio larger than 1:2) at $z\simeq 0.5-1.0$. By selecting mergers at this redshift interval, the merger remnants had time to relax before the end of the simulation at $z=0$. The stellar and halo mass at $z=0$ roughly match the values for the Milky Way, and so the merging galaxies are less massive than the Milky Way. We note, that the mergers are gas-rich, \new{which means that quenching is not important for the progenitors}.  

Major merger simulations with these four initial conditions were first presented in \citet{2016MNRAS.462.2418S} and \citet{2017MNRAS.470.3946S}. The new simulations from \citet{2020arXiv201113947W} improve upon these original simulations by including a magnetic field (ideal MHD) and tracer particles, which can be used to follow Lagrangian gas flows. Specifically, we track Lagrangian gas flows using Monte-Carlo tracers (\citealt{2013MNRAS.435.1426G}; we use 5 tracers per gas cell at the start of the simulation). Including a treatment of the magnetic field evolution is an important addition, because it affects the morphology of the remnant discs \citep{2020arXiv201113947W}, and it can suppress fluid instabilities \citep{2015MNRAS.449....2M,2019MNRAS.489.3368B,2020MNRAS.499.4261S}, including in the CGM \citep{2020arXiv200807537V}.

The physics underlying the galaxy formation model is identical to that used in the Auriga simulations \citep{2017MNRAS.467..179G}. Gas cells with a number density larger than the star formation threshold density ($n_\text{SFR}=0.13$ cm$^{-3}$) are described by a two-phase subgrid-model \citep{2003MNRAS.339..289S}, which takes into account a cold phase with a temperature $10^3$ K and an ambient hot phase. The equilibrium between the two phases \new{takes into account} radiative cooling and energy evaporated by supernovae. \new{Stellar population particles in our model} are stochastically spawned by cells with a density larger than $n_\text{SFR}$. The free parameters of the subgrid model are calibrated to reproduce the observed scaling-relation between gas surface density and SFR surface density for disc galaxies in the local Universe \citep{1989ApJ...344..685K}.

When stars are formed elements are created, and for each gas cell and stellar population particle we track the abundance of H, He, C, N, O, Ne, Mg, Si, and Fe. We use radiative cooling of gas cells and assume a uniform ultra-violet background following \citet{2009ApJ...703.1416F}. The Auriga model also includes an Eddington-limited Bondi-Hoyle-Lyttleton growth rate for the black hole \citep{1944MNRAS.104..273B,1952MNRAS.112..195B} in the centre of the galaxies. The simulations use a quasar feedback model, \new{where the energy injection in the radio mode is designed to balance the X-ray luminosity} \citep[following][]{2000MNRAS.311..346N}. For a full description of the black hole model and other aspects of the simulation model, see \citet{2017MNRAS.467..179G}. For the full details of the simulations, see \citet{2020arXiv201113947W}. We note that our simulations A, B, C and D refer to 1330-3M, 1349-3M, 1526-3M, and 1605-3M in the nomenclature of \citet{2020arXiv201113947W}. For an overview of cosmological galaxy formation simulations we point the reader to \citet{2015ARA&A..53...51S}, \citet{2017ARA&A..55...59N}, and \citet{2013MNRAS.436.3031V,2020NatRP...2...42V}.

\subsection{Ionization modelling}\label{IonisationModelling}

The ionization state of a selection of atoms is calculated following the framework of \citet{2018MNRAS.475.1160H}, which computes a grid of CLOUDY equilibrium models \citep{2013RMxAA..49..137F} determining the ion fraction as a function of density, gas metallicity and temperature. We assume the only ionizing radiation source to be the UV background following \citet{2009ApJ...703.1416F}. We include self-shielding of dense gas from the UV background following \citet{2013MNRAS.430.2427R}. The result of the calculation is a determination of the number density of e.g. \iona{H}{i}, \iona{O}{i}, \iona{O}{vi}, \iona{O}{vii}, \iona{O}{viii}, \iona{Si}{ii}, \iona{Si}{iii}, \iona{Si}{iv}, and \iona{Fe}{ii} for each gas cell. In this paper we focus on the \iona{H}{i} distribution.

In our subgrid multiphase model for the ISM, each star-forming gas cell has a contribution from cold star-forming clouds (with a temperature of $10^3$ K), which dominates the mass-budget, and hot gas surrounding them. In the ionization modelling, we \new{therefore} consider star-forming gas cells to be fully neutral.

\subsection{Definition of the CGM}

A \new{unique} definition of the CGM does not exist, but it is commonly thought of as the gas reservoir, where outflows from the ISM mix with cosmological inflows. For the purpose of this paper, we define the ISM and CGM regions based on the variable, $s \equiv \min(r_1/R_{200,1},r_2/R_{200,2})$, where $r_1$ denotes the distance of a gas cell from galaxy 1, and $R_{200,1}$ the radius \new{at which} the spherically averaged density of galaxy 1's halo equals 200 times the critical density of the Universe. For galaxy 2, the variables have a subscript of 2. $R_{200}$ lies in the range $210-249$ kpc for the merger remnants at $z=0$, and $142-187$ kpc at $z=0.69$, when the mergers are ongoing. We formulate the definitions of the ISM, CGM, and the gas outside the haloes as follows:
\begin{align*}
\text{We define, }& s \equiv \min(r_1/R_{200,1},r_2/R_{200,2}).\\
\textbf{ISM:}\quad& \text{If } s \leq 0.2\text{, a gas cell is assigned to the ISM.}\\
\textbf{CGM:}\quad& \text{If } 0.2 < s \leq 1.0 \text{, a gas cell is assigned to the CGM.}\\
\textbf{Outside halo:}\quad&\text{If } s>1.0 \text{ a gas cell is outside the merging haloes.}
\end{align*}
Each gas cell in a simulation fulfils exactly one of these definitions. For two merging galaxies, we sketch our definition of gas in the ISM, CGM and outside the halo in Fig.~\ref{ISM_CGM_Definition}.

With the above definitions, the CGM is bounded by two spherical surfaces. This choice has been made, because the volume of the CGM is usually dominated by warm gas ($\simeq 10^5-10^6$ K), which has a nearly spherical distribution with only a weak contribution from biconical outflows (see, e.g., \citealt{2016MNRAS.460.2157O,2018MNRAS.481..835O}).

These thresholds for defining the ISM and CGM are of course arbitrary, and we note that we have chosen quite conservative limits for the CGM region, because the disc and ISM of a galaxy normally only extend to 10-15 per cent of $R_{200}$ \citep{2014MNRAS.437.1750M}. These choices are, however, fully sufficient for our purpose, which is to demonstrate gas transfer to the bridge and ISM from the CGM and \new{outside} of the haloes.



\section{Results}\label{Results}

\subsection{Orbits and merger-induced star formation peaks}

\begin{figure}
\centering
\includegraphics[width = 1.0\linewidth]{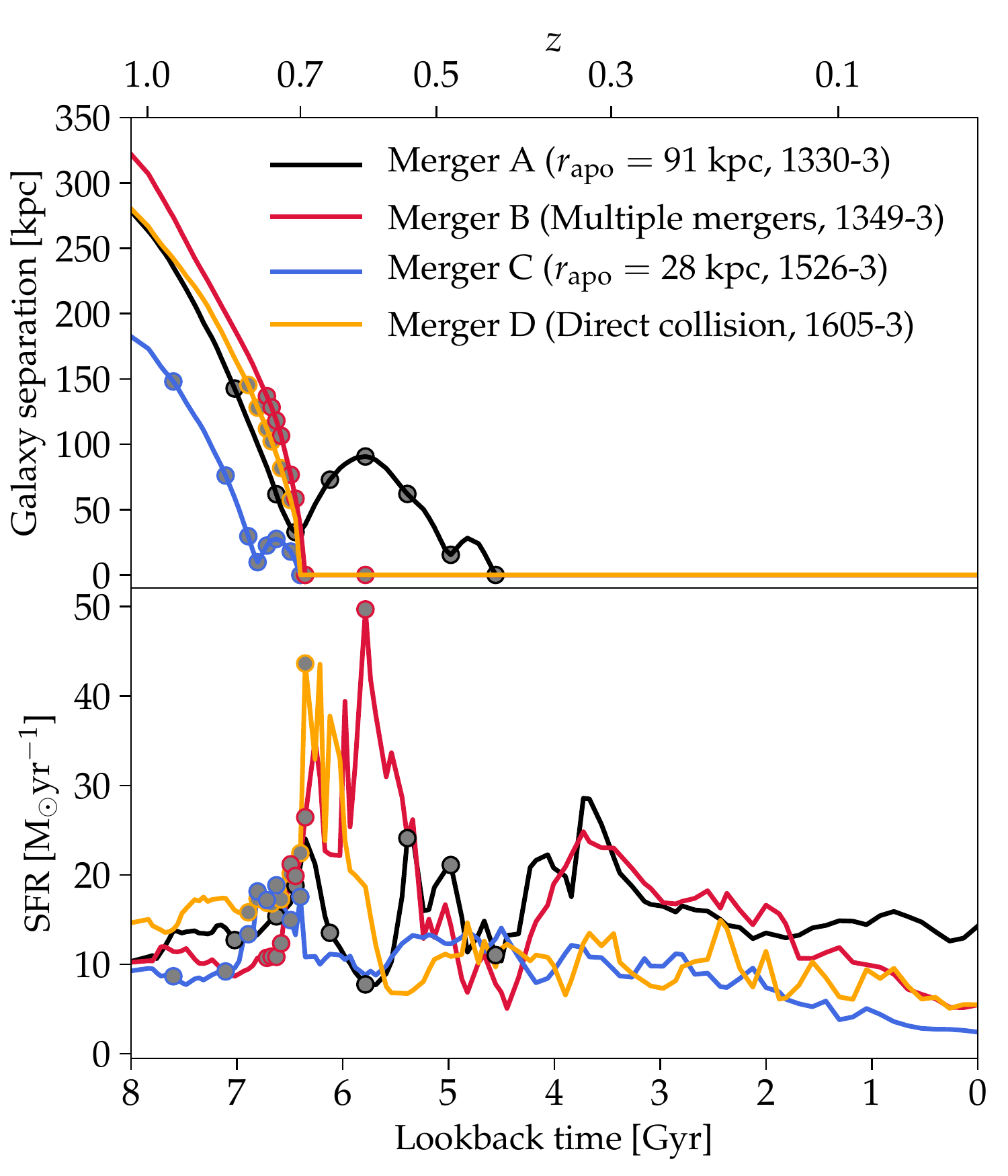}
\caption{The evolution of the distance between the two merging galaxies (\emph{upper panel}) and the star formation rate (\emph{lower panel}). Two mergers have multiple closest passages (A and C), and two mergers are direct collisions (B and D). In all cases, we see bursts of star formation at or shortly after the pericentric passages or the final coalescence. In the legend we have noted the apocentric separations, $r_\text{apo}$, for the mergers with multiple passages. The \emph{grey-filled circles} indicate instances in time, which will be studied further below.}
\label{MS_Orbits}
\end{figure}

To characterise the simulated galaxy mergers, we plot the distance between the two main progenitors and the SFR for each merger simulation in Fig.~\ref{MS_Orbits}. In the SFR calculation, we include the gas bound to the two main progenitor galaxies of the merger. All of the mergers have a close passage at $z\simeq 0.7$, and by $z\simeq 0.4$ they have all reached the final coalescence. Two mergers, B and D, are \emph{direct collisions}, and they merge at the first pericentric passage. The former is peculiar, because multiple galaxies participate in the merger. Both of the direct collision mergers have a violent starburst in comparison to the mergers with multiple pericentric passages (A and C). In simulation A the two merging galaxies orbit each other for approximately 2 Gyr before the final coalescence, and there is a correspondence with the two pericentric passages and a peak in the SFR. \new{We note that the SFR peaks approximately 1 Gyr after the identified time of coalescence. Visual inspection of images reveals that the late SFR is indeed caused by interaction of the two nuclei of the merging galaxies. This is because the galaxy nuclei continue to orbit each other after our halo finder has marked the two galaxies as merged.}

The overall properties of the merger orbits and star formation peaks are consistent with the analysis in \citet{2016MNRAS.462.2418S}, which has the same initial conditions but a slightly different physical model (see Sec.~\ref{SubSecSimulations}).

\subsection{Bridges in mergers}

The formation of a bridge between the two first passages is expected from the idealised modelling of stellar orbits in mergers \citep{1972ApJ...178..623T} and modern reconstructions of observed mergers \citep{2013ApJ...771..120P}. Studying the structure of the bridge is interesting, because it opens a possibility to study star formation in extreme environments outside of galaxy discs (we will return to this in Sec.~\ref{SecExtremeEnv}).

In idealised closed-box models of galaxy mergers, it is evident that the bridge is formed from stripped gas (or outflows) from the galaxy discs, because gas accretion and gas in the CGM are not included in such setups. In a cosmological simulation, the gaseous component of a bridge may, however, be more complicated, because of the potential contribution of gas accreted from the CGM and the cosmological surroundings.

\begin{figure*}
\centering
\includegraphics[width = 1.0\linewidth]{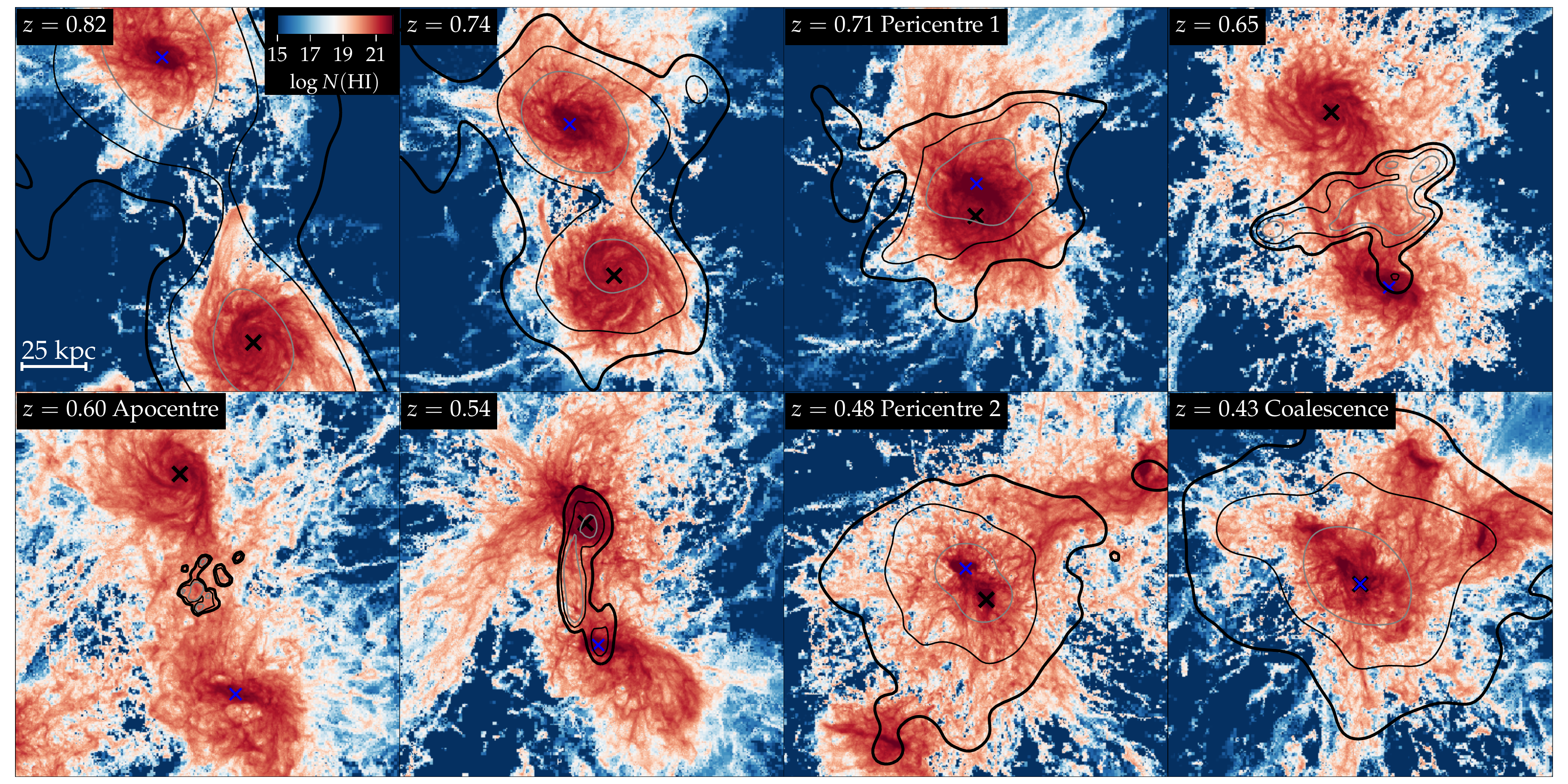}
\caption{An image sequence showing the evolution of the \iona{H}{i} column density for the merger in simulation A ($r_\text{apo}=91$ kpc). The x-symbols show the centres of the two main galaxies participating in the merger. Between the first and second passage (see panels from $z=0.65$ to $z=0.54$) a bridge of dense \iona{H}{i} gas arises between the galaxy discs. The contours encompass 50, 75, and 90 per cent of the tracer particles selected to be in the galaxy bridge at $z=0.60$; the gas in the bridge is stripped from the galaxy discs as well as accreted from the CGM.}
\label{Fig100_HI_sequence_1330-3MHD}
\end{figure*}

\begin{figure*}
\centering
\includegraphics[width = 1.0\linewidth]{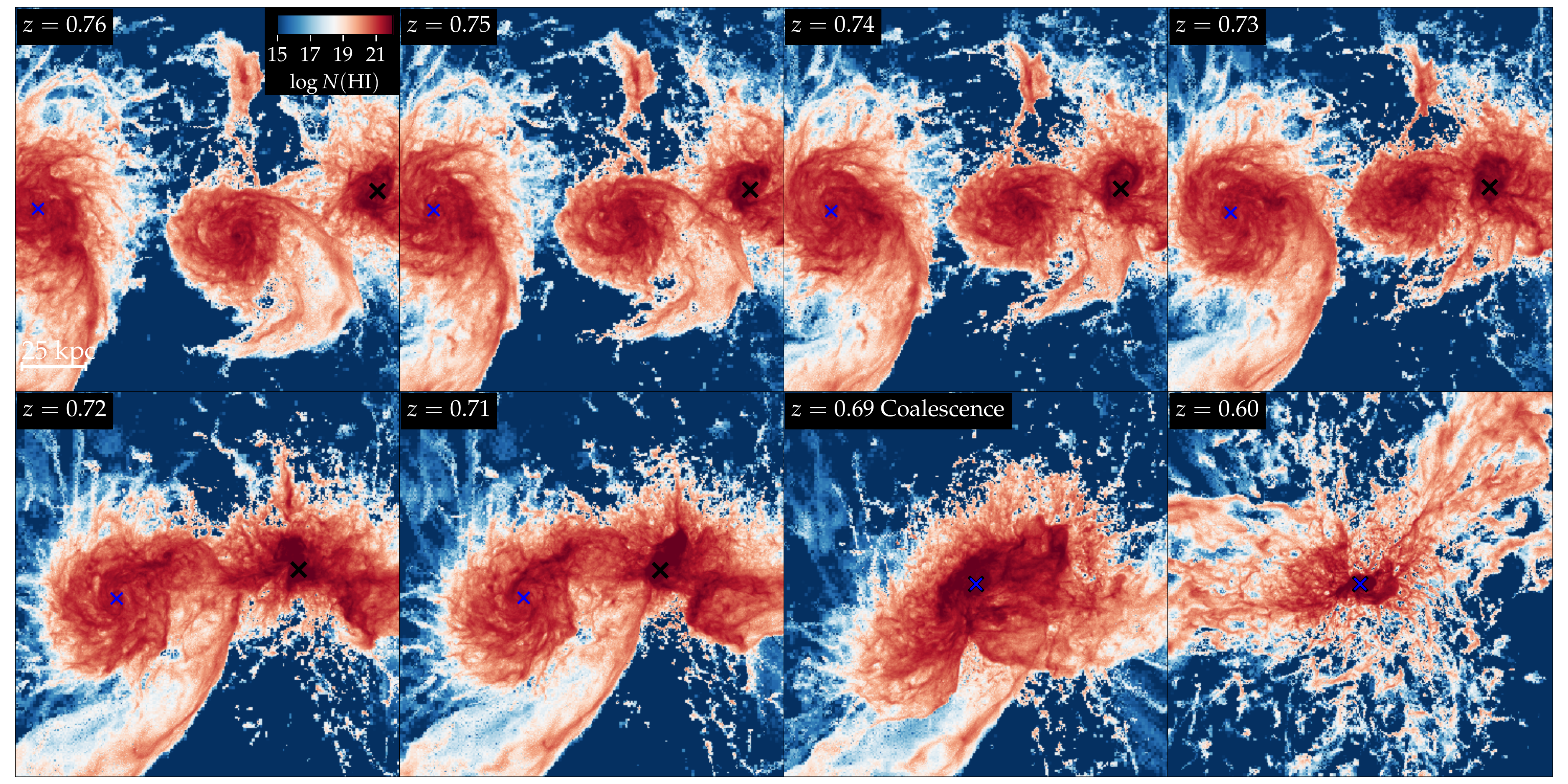}
\caption{Multiple galaxies participate in the merger in simulation B. The secondary progenitor, which is marked with a blue x-symbol, reveals a visible tail, especially at $0.73\leq z\leq 0.75$, which arises from the gravitational forces during the merger. Detailed inspection reveals complex gas flows caused by the multiple galaxies participating in this merger. We do not identify a well-defined bridge in this simulation, so we do not show any contours of gas tracers from the bridge.}
\label{Fig100_HI_sequence_1349-3MHD}
\end{figure*}

\begin{figure*}
\centering
\includegraphics[width = 1.0\linewidth]{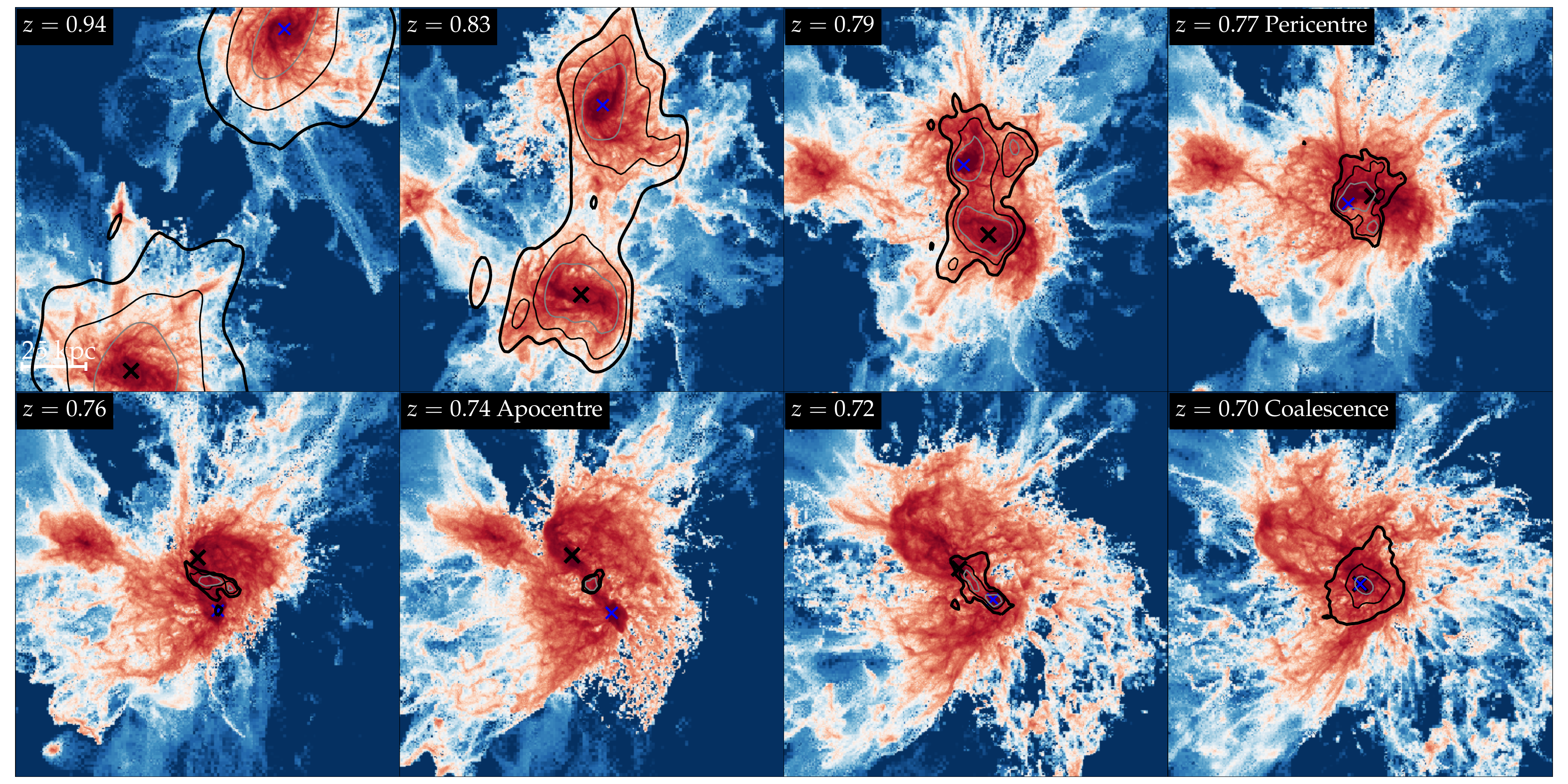}
\caption{The merger in simulation C has an impact parameter ($r_\text{apo}$) of 28 kpc, and as for simulation A we see a bridge forming \new{at the apocentre} at $z=0.74$. Here we select gas tracers to reveal the origin of the bridge. The majority of the gas comes directly from the discs of the two main progenitors.}
\label{Fig100_HI_sequence_1526-3MHD}
\end{figure*}

\begin{figure*}
\centering
\includegraphics[width = 1.0\linewidth]{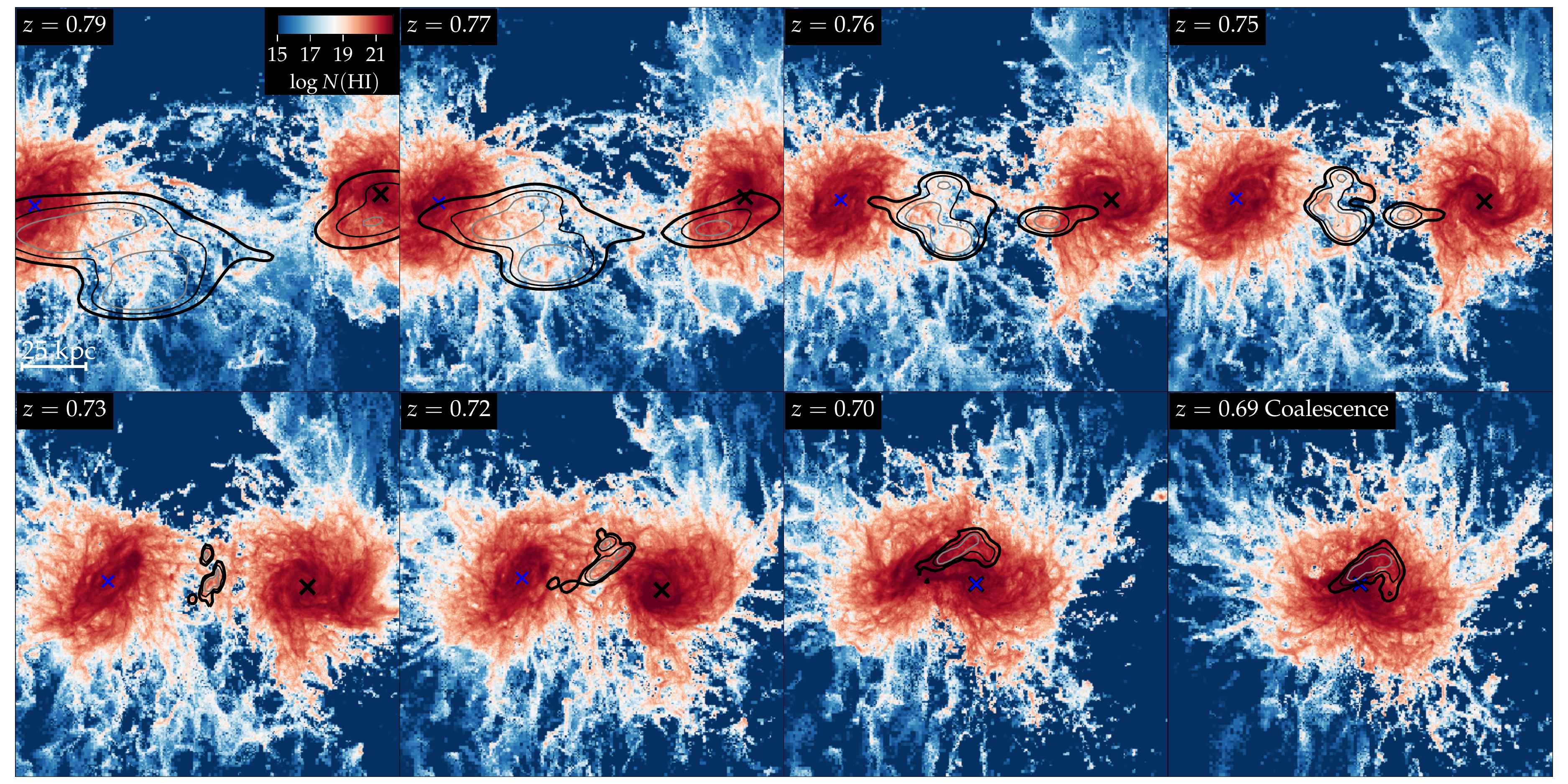}
\caption{Simulation D is a direct collision merger, implying that there is no second passage. Therefore, a long-lived bridge as remarkable as in simulation A and C never emerges, but \new{signs of dense gas resembling a bridge are seen} at $z=0.73$. Gas tracers in this bridge show that it forms from gas stripped from the discs and from cosmological inflows.}
\label{Fig100_HI_sequence_1605-3MHD}
\end{figure*}

To identify gas in galaxy bridges in our simulations, we project the \iona{H}{i} column density of the four mergers in Fig.~\ref{Fig100_HI_sequence_1330-3MHD}-\ref{Fig100_HI_sequence_1605-3MHD}. The panels have a width of 150 kpc, and we use a random direction of the projection (along the $z$-axis in the simulations). The times plotted in each panel are marked with circles in Fig.~\ref{MS_Orbits}. In the case of mergers with non-zero impact parameters (simulations A and C) we specifically include the \new{pericentric passage and the apocentre}.

\begin{figure*}
\includegraphics[width=\linewidth]{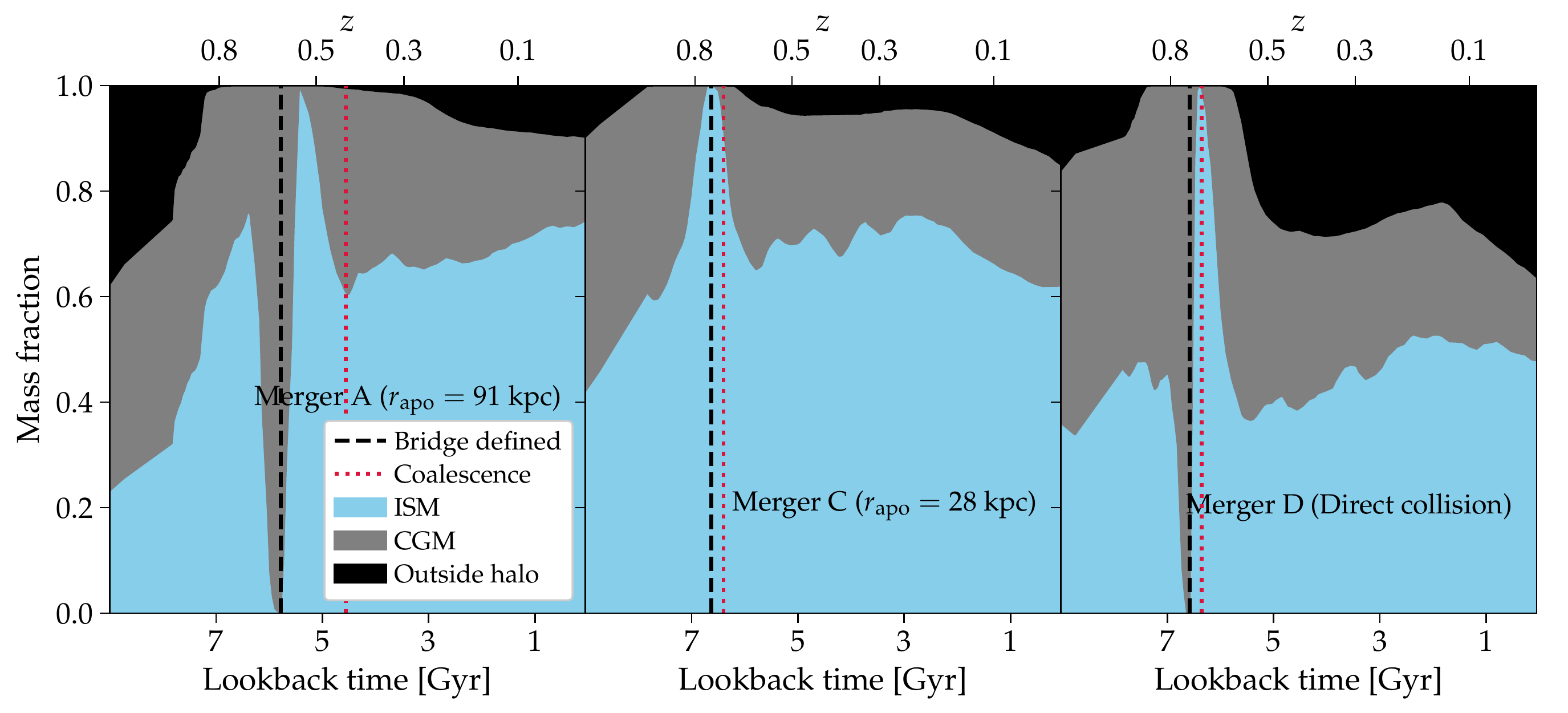}
\caption{We trace the gas residing in the bridge at the time marked by the dashed vertical line. These are the same gas reservoirs as tracked in Fig.~\ref{Fig100_HI_sequence_1330-3MHD}, \ref{Fig100_HI_sequence_1526-3MHD} and \ref{Fig100_HI_sequence_1605-3MHD}. Each tracer particle is reassigned to either the ISM, CGM or outside the halo as a function of time, and this allows us to determine the contribution of the three gas reservoirs in building the bridge. The gas constituting the bridge has a significant contribution from CGM and ISM gas: 1 Gyr before the bridge was most prevalent, 47--67 per cent of the gas was in the ISM, and 33-48 per cent was in the CGM. Shortly \new{after the apocentre} the gas constituting the bridge enters the ISM of the two galaxies ($\gtrsim 98$ per cent), and at later times significant fractions enter the CGM or leaves the haloes completely.}
\label{Fig100_HI_sequence_TracedGasInBridge}
\end{figure*}

\subsubsection{Bridges after first coalescence (simulation A and C)}\label{SimAandCSubSec}

We start by characterising simulation A and C, which have multiple passages. At the \new{apocentre}, these simulations reveal a bridge connecting the galaxy discs (see Fig.~\ref{Fig100_HI_sequence_1330-3MHD} and Fig.~\ref{Fig100_HI_sequence_1526-3MHD}, respectively). It is expected, but still reassuring, to see the formation of a bridge during a merger in a fully self-consistent cosmological galaxy formation simulation. To reveal the origin of the gas in the bridge we select gas cells in the bridge \new{at the apocentre}, where the bridge appears most prevalent, and track the gas flows using \emph{tracer particles}.

We decompose the coordinates of each gas cell into a perpendicular and parallel component of the vector connecting the two merging galaxies. The former is defined as,
\begin{align}
d_\perp \equiv \frac{|(\pmb{x}_2-\pmb{x}_1)\pmb{\times} (\pmb{x}_\text{gas}-\pmb{x}_1)|}{  |\pmb{x}_2-\pmb{x}_1| },
\end{align}
where $\pmb{x}_1$ and $\pmb{x}_2$ are the coordinates of the primary and secondary galaxy in the merger, respectively, and $\pmb{x}_\text{gas}$ is the coordinate of a gas cell. This quantity is unsigned and it measures the absolute distance from each gas cell to the line, which passes through $\pmb{x}_1$ and $\pmb{x}_2$. We define the parallel component as,
\begin{align}
d_\parallel \equiv \frac{(\pmb{x}_2-\pmb{x}_1)\pmb{\cdot} (\pmb{x}_\text{gas}-\pmb{x}_1)}{  |\pmb{x}_2-\pmb{x}_1| }.
\end{align}
This quantity is signed, and gas cells between the merging galaxies have $0 < d_\parallel < |\pmb{x}_2-\pmb{x}_1| $.

With these quantities, we can select tracers residing in gas cells in the bridge. For simulation A, we select the tracers at $z=0.60$, which is the time \new{of the apocentre}. We select tracers, which satisfy the following two criteria:
\begin{itemize}
\item Gas cells should be located between the galaxies. We specifically require $0.4 < d_\parallel / |\pmb{x}_2-\pmb{x}_1| < 0.6  $ and $d_\perp < 20 $ kpc.
\item We furthermore only select dense gas with a number density of $n\geq 0.13$ cm$^{-3}$.
\end{itemize}
We visualise the tracers by showing contours encapsulating 50, 75 and 90 per cent of their distribution (Fig.~\ref{Fig100_HI_sequence_1330-3MHD}). Naturally, we see the most concentrated distribution at \new{the apocentre}, where the tracers are selected. Prior to this, at $z=0.74$, these gas tracers are mostly residing in the two galaxy discs, but there is also a visible contribution from outside the disc region (we quantify the mass fractions in Sec.~\ref{MassBudget}). At coalescence, tracers are abundant in the central parts of the merger remnant and outside the galaxy.

For simulation C we also select tracers from the bridge \new{at the apocentre} ($z=0.74$). The colliding galaxies here have a smaller separation, so we adjust the cutoff of the perpendicular distance to $d_\perp < 10 $ kpc -- except for this we use the same selection criteria as for simulation A. Prior to the merger the majority of the gas again comes from the discs (see Fig.~\ref{Fig100_HI_sequence_1526-3MHD}), but there is also a visible contribution from a satellite galaxy (see the panels showing $z=0.94$ and $z=0.83$).

\subsubsection{The direct collisions (simulation B and D)}

In simulation B the merger's two main progenitors undergo a direct collision. In Fig.~\ref{Fig100_HI_sequence_1349-3MHD} we show \iona{H}{i} images, which reveal more than two galaxies to participate in the merger. \new{We note the existence of a tail forming in one of the merging galaxies (seen below the galaxy marked with a blue cross at $z=0.75$ to $z=0.73$), and at $z=0.72$ we see that the two main galaxies are connected with a gaseous bridge-like structure. The field of view is, however, influenced by other galaxies than the two main progenitor galaxies, so we do not perform our tracer analysis for this simulation}. The fact that it is a direct collision could also be the reason why a bridge is not appearing, because a bridge is usually most prevalent after the first passage.

 We note that the existence of such a merger with $\geq 3$ galaxies interacting exemplifies how important the cosmological surroundings of a merger can be.

Simulation D, which is also a direct collision, is shown in Fig.~\ref{Fig100_HI_sequence_1605-3MHD}. This is a fairly clean merger, with only two massive galaxies participating. Again, a prominent long-lived bridge does not arise for the reasons stated above. There are, however, still signs of dense gas between the galaxies, especially at $z=0.73$. \new{This feature could, for example, be caused by tidal compression (as seen in the simulations of \citealt{2009ApJ...706...67R}), shocks or ram pressure.} Here we select gas tracers using the same selection criteria as in simulation A (except for the different redshift). By examining the image at $z=0.79$ \new{we determine that this gas consists} partially of stripped gas from the two merging galaxies as well as inflows from the CGM. At coalescence this gas reservoir resides in the inner parts of the merger remnant.

\subsubsection{The role of tidal stripping from the disc}

\new{Tidal stripping of material from a galaxy's disc is the \emph{standard mechanism} \citep{1972ApJ...178..623T} for forming a stellar bridge. If a bridge is present in both \iona{H}{i} column density and stellar surface density, the origin of the bridge is most likely tidal stripping. On the other hand, if a stellar bridge is absent and a \iona{H}{i} bridge is present, gaseous processes (such as ram pressure stripping, shocks and gas accretion) must be causing the formation of the bridge. We will refer to such gaseous processes as \emph{collisional processes}. In Appendix~\ref{StellarMaps} (Fig.~\ref{Fig104_Stars_sequence_1330-3MHD}--\ref{Fig104_Stars_sequence_1605-3MHD}) we plot stellar surface density maps, which can be directly compared to our \iona{H}{i} maps.}

\new{For the direct collision merger D, the gaseous bridge origins from collisional processes, because a stellar bridge is absent. The same is the case for merger A at the time of the apocentre. Merger C has both a remarkable stellar and gaseous bridge at the apocentre, so here tidal stripping is the dominant process. For an in-depth analysis, see Appendix~\ref{StellarMaps}.}

\new{In \citealt{2018MNRAS.480.3069P} (see their fig. 3, and the associated text) it was impossible to explain the observed \iona{H}{i} bridge in between two merging galaxies (NGC 4490 and 4485) using purely collisionless modelling (following the framework of \citealt{2009AJ....137.3071B}). The authors speculated that the absence of the bridge in their modelling was mainly caused by the lack of hydrodynamics. Although it remains to be seen whether this is indeed the reason for them not being able to reproduce the bridge, our simulations show that mergers can have a much more prevalent \iona{H}{i}-bridge in comparison to the stellar bridge.}

\subsection{Quantifying the mass budget of the gas in bridges}\label{MassBudget}


In Fig.~\ref{Fig100_HI_sequence_TracedGasInBridge} we show the evolution of the mass fraction of gas selected to be in the bridge. Simulation B did not reveal a bridge in our previous analysis, so we only carry out this analysis for the other three simulations. The time when we select the tracers in the bridge is shown by a \emph{dashed vertical line}. When the bridge is most prevalent its gas is either characterised as CGM (for merger A and D), or it is a part of the ISM in the case of merger C, where the apocentric separation is low enough for the bridge-region to be within $0.2 R_{200}$ of one of the merging galaxies. 1 Gyr prior to this 47--67 per cent of the gas was in the ISM, and 33--48 per cent is accreted from the CGM (and 0--5 per cent comes from outside the haloes). The gas bridge is therefore not simply consisting of stripped material from the discs, which would be the simplest and most intuitive Ansatz, but rather a mix of gas from our three different gas reservoirs.

Shortly after the merger (i.e. within $\lesssim 1$ Gyr after the coalescence), the bridge gas has entered the ISM in all simulations, and a few Gyr later fractions of it have been ejected to the CGM or out of the halo, as expected from feedback processes (such as stellar winds, supernovae, cosmic rays and black holes) in the contemporary galaxy formation paradigm.

\begin{figure*}
\includegraphics[width=\linewidth]{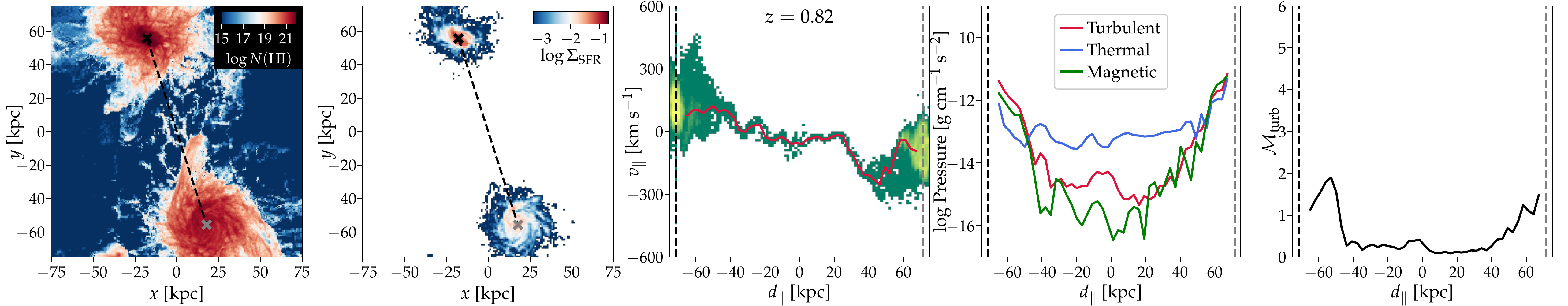}
\includegraphics[width=\linewidth]{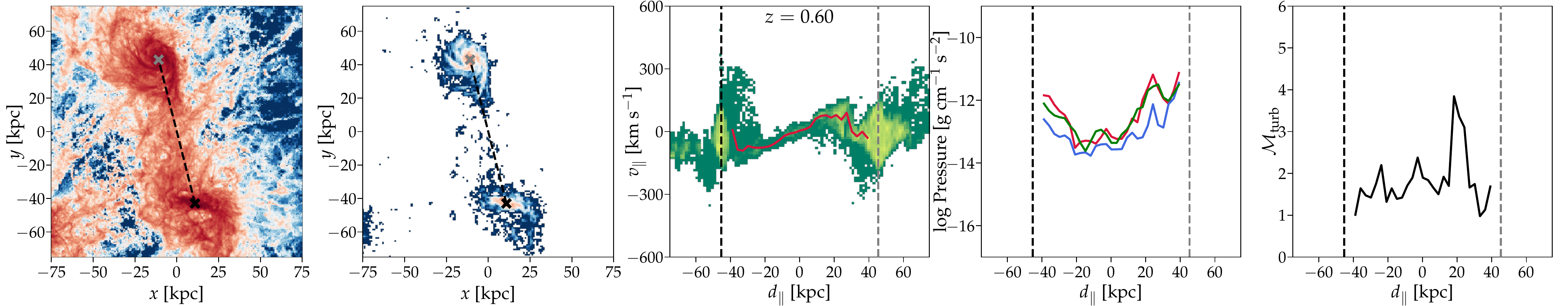}
\caption{For merger A we show the \ion{H}{i} column density (measured in cm$^{-2}$), star formation rate column density ($\Sigma_\text{SFR}$, here measured in M$_\odot$ yr$^{-1}$ kpc$^{-2}$), velocity structure, pressure (turbulent, magnetic and thermal), and the Mach number of the turbulent velocity component. In the spatial maps we mark the location of the merging galaxies with x-symbols, and in the panels with $d_\parallel$ as abscissa we mark the galaxies with a vertical dashed line (with a matching colour). The upper panels show a time well before the first pericentric passage, and the lower panels show the (first) apocentre.}\label{Fig100_HI_sequence_BridgeContractionAfterReferee_SinglePanel1330-3_1330-3MHD}
\end{figure*}

\begin{figure*}
\includegraphics[width=\linewidth]{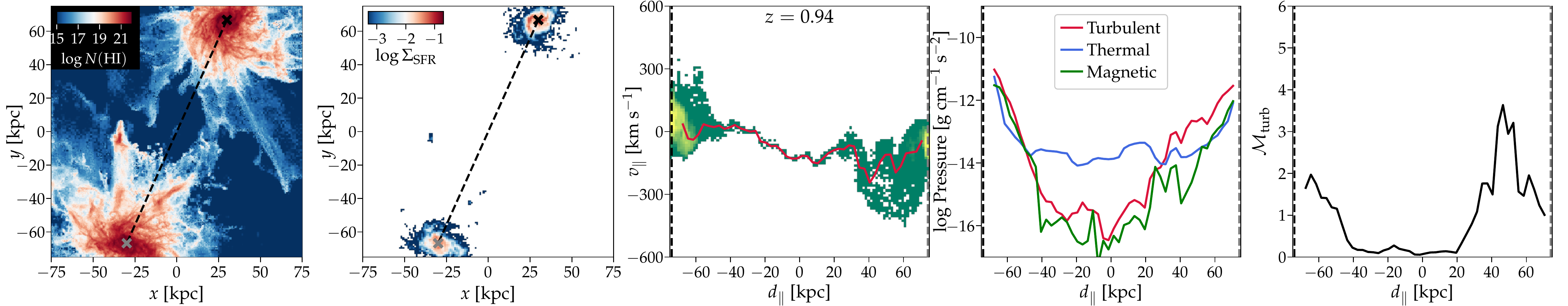}
\includegraphics[width=\linewidth]{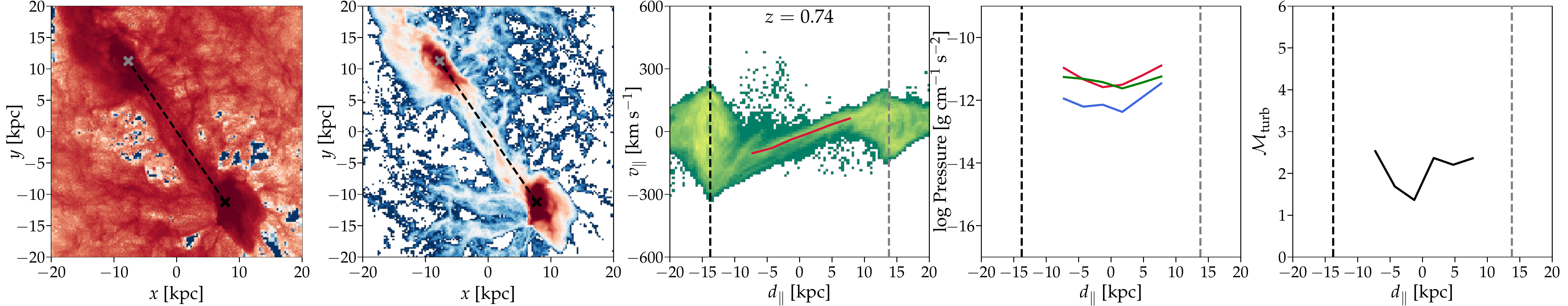}
\caption{Same as Fig.~\ref{Fig100_HI_sequence_BridgeContractionAfterReferee_SinglePanel1330-3_1330-3MHD}, but for merger C.}\label{Fig100_HI_sequence_BridgeContractionAfterReferee_SinglePanel1526-3_1526-3MHD}
\end{figure*}

\begin{figure*}
\includegraphics[width=\linewidth]{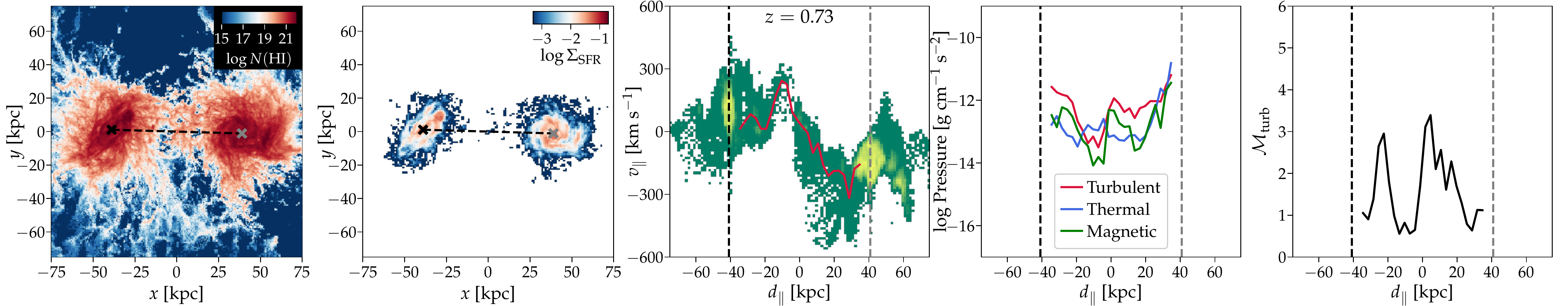}
\caption{Same as Fig.~\ref{Fig100_HI_sequence_BridgeContractionAfterReferee_SinglePanel1330-3_1330-3MHD}, but for merger D at the time where a bridge is most visible (prior to coalescence).}\label{Fig100_HI_sequence_BridgeContractionAfterReferee_SinglePanel1605-3_1605-3MHD_063}
\end{figure*}

\subsection{Star formation and turbulence under extreme conditions in the bridge}\label{SecExtremeEnv}

The gas in the bridge of a galaxy merger is useful for probing star formation in extreme environments outside of galaxy discs. \citet{2015MNRAS.446.2038R}, for example, identified star formation in the bridge in between two simulated merging galaxies. Observations of the merging \emph{Taffy galaxies} reveal a bridge of molecular gas \citep{2003A&A...408L..13B} with a low star formation efficiency \citep{2012A&A...547A..39V,2021A&A...647A.138V}. This could be explained by \new{supersonic turbulence moderating star formation} \citep[as seen in turbulence models, e.g.][]{2012ApJ...759L..27P,2012ApJ...761..156F}. In the following we analyse the role of turbulence in the bridge of our simulations.

We characterise the conditions in the bridges in Fig.~\ref{Fig100_HI_sequence_BridgeContractionAfterReferee_SinglePanel1330-3_1330-3MHD}, \ref{Fig100_HI_sequence_BridgeContractionAfterReferee_SinglePanel1526-3_1526-3MHD} and \ref{Fig100_HI_sequence_BridgeContractionAfterReferee_SinglePanel1605-3_1605-3MHD_063}. We show the projected \ion{H}{i} column density and the SFR surface density to visualize the mergers. \new{We note that our subgrid star formation model is tuned to reproduce an ISM characteristics of disc galaxies. This means our model does not include potential effects that may impact the star formation law in the bridge}.

\new{For further quantitative analysis, we select gas cells within a distance of $d_\perp\leq 3$ kpc from the line connecting the two merging galaxies. We create bins with width 3 kpc along the $d_\parallel$-axis. The first bin's coordinate centre is 5 kpc from the main progenitor, and last bin's centre is 5 kpc before reaching the secondary progenitor. With these choices we obtain cylindrical bins with volumes of $\pi \times (3 \text{ kpc})^3 \simeq 85$ kpc$^3$ and a linear size of 4.4 kpc (calculated as the volume to the power of $1/3$). Such large volumes correspond to a length scale larger than giant molecular clouds (GMCs), so our analysis does not enable us to study turbulence at the GMC scale. Nevertheless, it still enables a determination of the turbulence on larger scales.}

We plot the velocity component along the connecting axis ($v_\parallel$), the pressure (thermal, turbulent and magnetic), and the turbulent Mach number. \new{We define the turbulent pressure by calculating the velocity dispersion ($\sigma_v$) for the cells in each cylindrical bin (i.e. in the rest velocity frame for each bin). The turbulent pressure is then calculated as $\frac{1}{2} \rho \sigma_v^2$. In each cylindrical bin, we determine the turbulent Mach number, $\mathcal{M}_\text{turb}\equiv \sigma_v/c_s$ ($c_s$ is the sound speed), by first calculating averages of the kinetic end thermal energy (per unit mass) by mass-weighing $\sigma_v^2$ and $c_s^2$, and subsequently we extract $\mathcal{M}_\text{turb}$ from their ratio.}

\subsubsection{The mergers with multiple pericentric passages}

The state of the gaseous bridge for the mergers with multiple passages is shown in Fig.~\ref{Fig100_HI_sequence_BridgeContractionAfterReferee_SinglePanel1330-3_1330-3MHD} (simulation A) and Fig.~\ref{Fig100_HI_sequence_BridgeContractionAfterReferee_SinglePanel1526-3_1526-3MHD} (simulation C). We show the redshift of \new{the first apocentre} (lower panels) and a time well before the first pericentric passage (upper panels). As we saw in Sec.~\ref{SimAandCSubSec} the bridge is more pronounced \new{at the apocentre} in comparison to before the first pericentric passage. In these figures we additionally show that star-forming clumps exist in the bridge both in simulation A and C \new{at the apocentre}.

In the plot of $(d_\parallel,v_\parallel)$ we see a negative gradient in between the two galaxies at the earliest shown redshift, and a positive gradient is seen \new{at the apocentre}. This means that gas is \new{flowing towards and away from} the centre of the bridge (where $d_\parallel=0$), respectively. \new{At the earliest shown redshift, the thermal pressure dominates in the bridge region, but at later times the turbulent and magnetic pressures reach approximate equipartition as a result of the merging process. At the same time supersonic turbulence \new{($\mathcal{M}_\text{turb}\simeq 2.0$ and $1.6$ for simulation A and C, respectively)} is seen in the centre of the bridge (again, at $d_\parallel=0$). For the mergers with multiple passages we indeed see extreme physical conditions in the bridge \new{at the apocentre}, and this highlights the fact that mergers are useful laboratories for star formation!}

\subsubsection{Direct collisions -- and the Taffy galaxies}

In Fig.~\ref{Fig100_HI_sequence_BridgeContractionAfterReferee_SinglePanel1605-3_1605-3MHD_063} we show merger D, which is a direct collision. We show a redshift of $z=0.73$, which is slightly before the coalescence, when the bridge is visually most pronounced (following the \ion{H}{i} panels in Fig.~\ref{Fig100_HI_sequence_1605-3MHD}). The bridge shares some of the characteristics seen for the mergers with multiple passages. Star formation occurs in the bridge, and the velocity structure shows that gas is \new{flowing towards the central point}. \new{At $0\geq d_\parallel \geq 10$ kpc, we identify supersonic turbulence with $\mathcal{M}_\text{turb}\simeq 3.3$, which is high in comparison to the analysed epochs for the mergers with multiple passages.}

The Taffy galaxies are observed before the first coalescence, and their bridges share many properties of our merger D. This simulation supports the scenario, where the gas in the bridge is supersonic turbulent. The star-forming clouds in our simulations are not resolved, but instead treated with a subgrid model, which takes into account the gas density, but does not explicitly depend on the velocity structure around the gas cells. Therefore, our star formation model is insufficient to address, whether and how the turbulence affects the star formation efficiency within clouds. But our simulations still reliably predict the presence of supersonic turbulence in the bridge.

\new{Using idealized non-cosmological simulations, it would be possible to achieve sufficient spatial resolution to resolve the turbulence at GMC scales (e.g., as in \citealt{2021arXiv210910356L}). In combination with a more sophisticated ISM model, which explicitly includes GMC formation, such simulations could reveal how our identified large-scale turbulence in the bridge influences star formation. A limitation of such an analysis would be that cosmological gas accretion is omitted.}

Simulation B also contains two galaxies undergoing a head-on collision, but we omit this from our analysis, because more than two galaxies are participating in this merger.

\begin{figure}
\centering
\includegraphics[width = 1.0\linewidth]{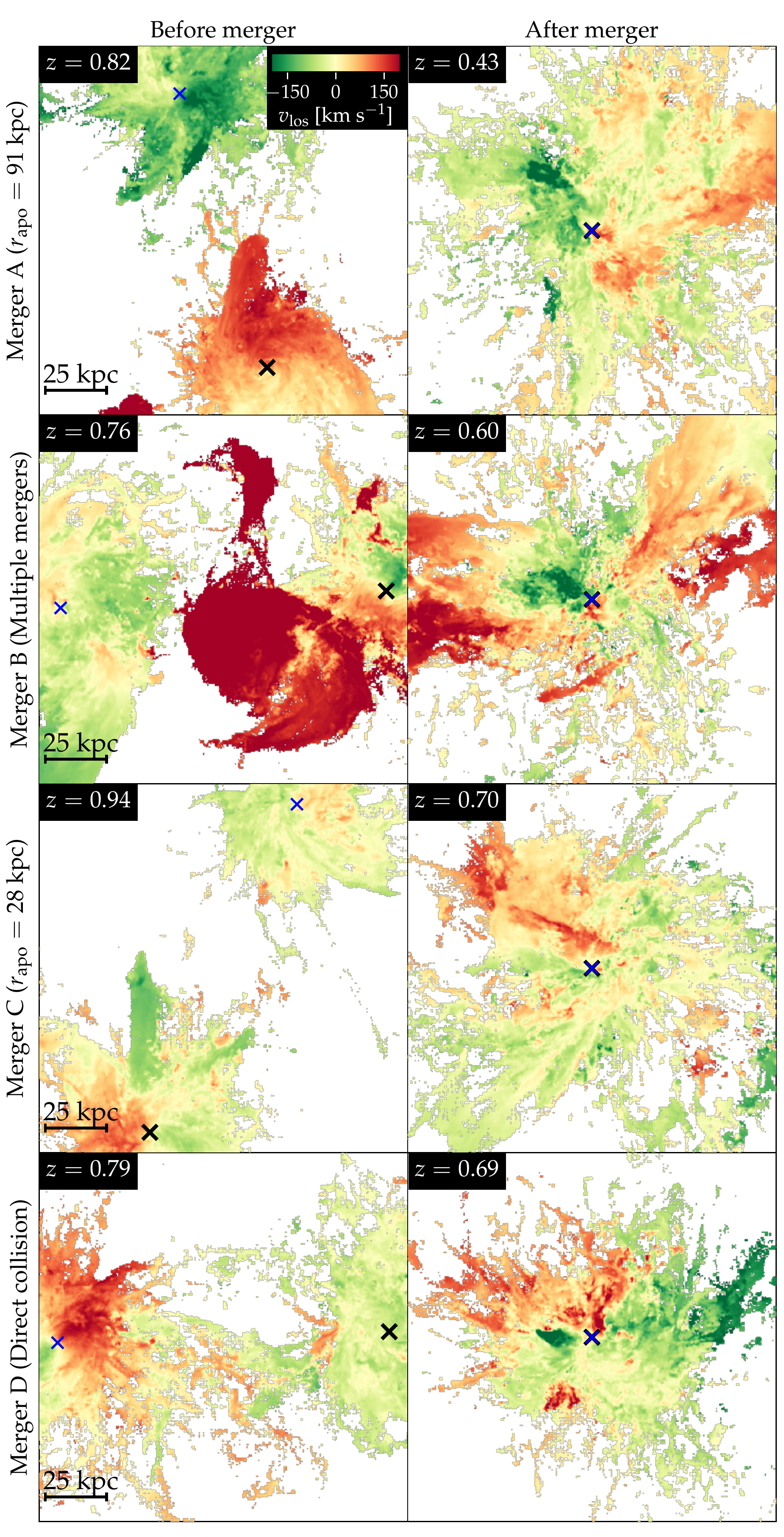}
\caption{\iona{H}{i} mass-averaged velocity projection maps before (left panels) and after (right panels) the merger for each simulation. Before the merger the progenitor galaxies have a line-of-sight velocity profile characteristic of a rotating disc galaxy. After the merger, the signature of rotation is weakened and replaced with a map revealing complex patterns characteristic of stripped gas.}
\label{Fig111_HIVelz}
\end{figure}

\begin{figure*}
\includegraphics[width = 0.95\linewidth]{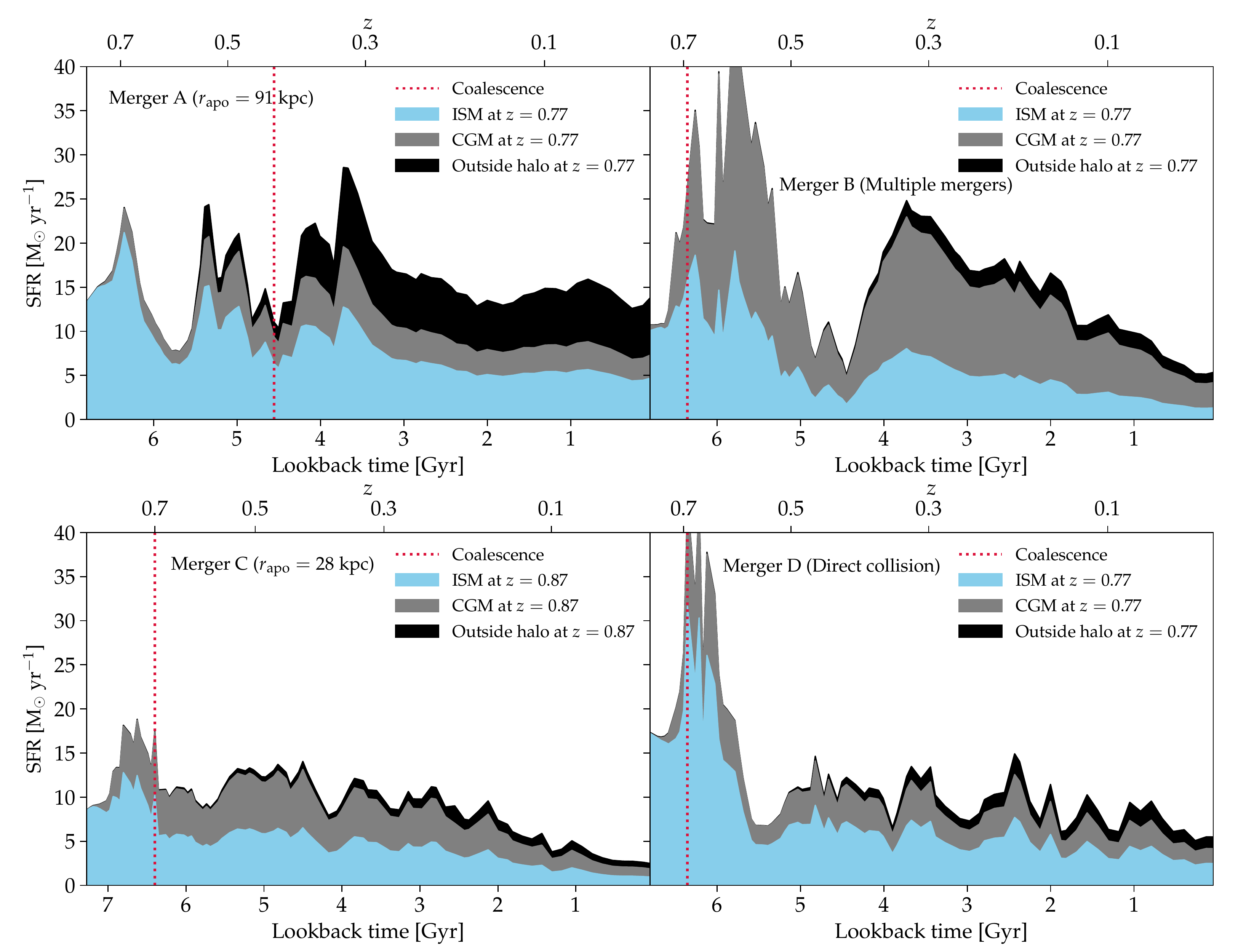}
\caption{Before the merger (at the redshift indicated in the legends) we tag gas as residing in the ISM, CGM, and outside the haloes, using the definitions from Fig.~\ref{ISM_CGM_Definition}. In each panel the largest shown lookback time is when we do this tagging. \new{We then track the SFR from each gas reservoir as a function of time. At the time of coalescence, a significant fraction of the SFR comes from gas that was in the CGM before the merger (27-51 per cent). At $z=0$, well after the merger, 24-45 per cent of the SFR occurs in gas that} was in the ISM before the merger, whereas 19-53 per cent of the SFR is fuelled by gas accreted from the CGM, and 22-48 per cent from outside the halo. Neither a CGM nor cosmological surroundings of galaxy mergers are included in traditional idealised simulations of galaxy mergers, so these significantly underestimate the SFR of the merger remnants.}
\label{MS_PlotTracedSFR}
\end{figure*}

%

\subsection{Gas kinematics}

Kinematic maps showing the gas velocity distribution are a powerful tool to constrain galaxy mergers. This was, for example, demonstrated by \citet{2018ApJ...868..112R} who studied the origin of gas around the \emph{Whale galaxy} (NGC 4631) with the Green Bank Telescope (with data from the HALOGAS project, see \citealt{2011A&A...526A.118H}). \citet{1994A&A...285..833R} used similar maps to study the gas around the same galaxy.

Our previously shown \iona{H}{i} maps revealed that gas was stripped and ejected from the galaxies during the merger. To reveal how a merger can affect the gas kinematics we study the \iona{H}{i} mass-averaged line-of-sight velocity ($v_\text{los}$) in Fig.~\ref{Fig111_HIVelz}. We only show pixels with $N(\text{\iona{H}{i}})> 10^{17}$ cm$^{-2}$, so we filter out low-density regions. Before the merger we generally see well-behaved profiles revealing the \iona{H}{i} gas to be rotating in discs. After the merger the distribution is much more complex, and clumpy velocity patterns are seen.

The difference between before and after the merger is most striking for merger A. Prior to the merger the gas with a (projected) distance smaller than 30 kpc follows a dipole pattern indicating that large-scale gas motions are correlated with the rotation of the galaxy discs. After the merger there are multiple gas clumps within 10 kpc of the merger remnant's centre, causing a transformation from a dipole pattern to unstructured gas motions. These are caused by gas stripping and (magneto-)hydrodynamical instabilities. It is therefore not surprising that \citet{2020ApJ...893L...3Z} observed signatures of shocks in their sample of merging galaxies.

Reproducing the \ion{H}{i} features connecting the Whale galaxy and a neighbouring galaxy, NGC 4656, at a projected distance of 100 kpc has proven a theoretical challenge. In between the galaxies are two observed \emph{spurs} connecting them (see spur 1 and 2 in fig. 4 from \citealt{2015AJ....150..116M} and fig. 1 from \citealt{2018ApJ...868..112R}). It has not been possible to reproduce these spurs, by simply modelling the stellar components of these galaxies \citep{1978A&A....65...47C,2015AJ....150..116M}. A possible reason is that it requires a full hydrodynamical treatment of the gas to reproduce the \iona{H}{i} features, as hydrodynamical effects (gas stripping and instabilities) are not captured by the modelling of collisionless stellar orbits. From a qualitative point of view the bridge connecting the two galaxies in simulation A at $0.54\leq z \leq0.68$ has a similarity with spur 1 and 2 seen between NGC 4631 and NGC 4656; noteworthy is the continuous distribution of \iona{H}{i} with $N_\text{\iona{H}{i}}\gtrsim 10^{19}$ cm$^{-2}$ filling the area between the two merging galaxies (compare Fig.~\ref{Fig100_HI_sequence_1330-3MHD} in our paper and fig.~1, upper panel, in \citealt{2018ApJ...868..112R}). By performing an idealised simulation of NGC 4627, 4631, and 4656 (i.e., a modern version of \citealt{1978A&A....65...47C}) using magneto-hydrodynamics, it would likely be possible to settle whether the lack of success in reproducing the spurs is caused by not including gas in previous simulations.

\subsection{Accretion of gas from the circumgalactic medium}

We now turn to studying the contribution of pre-merger CGM gas to the merger-induced SFR budget. We tag each gas cell well before the merger as belonging to the ISM, CGM, or as being outside the haloes of the two merging galaxies. We again use the classification introduced in Fig.~\ref{ISM_CGM_Definition}. For simulation A, B, and D we tag each gas cell following these definitions at $z=0.77$, and for simulation C we pick an earlier time of $z=0.87$, as this merger happens at a slightly higher redshift compared to the other simulations.

We track the gas in time and compute the reservoir's contributions to the two merging galaxies (see Fig.~\ref{MS_PlotTracedSFR}). When we tag the gas cells, nearly all the star formation happens in the inner regions of the two main progenitors (the ISM). An exception is simulation B, where more than two galaxies are merging, so there is a minor contribution from gas outside the two main progenitors. 

At the time of coalescence, 27-51 per cent of the SFR originates from gas accreted from the CGM. This means that during the merger gas is actively accreted from the CGM to the ISM. In the merger remnants at $z=0$, we see that 19-53 per cent of the SFR occurs in gas that was in the CGM before the merger, and only 24-45 per cent of the SFR comes from gas that was in the ISM prior to the merger. The CGM therefore plays a key role in fuelling the merger remnants and reigniting star formation.

\begin{figure*}
\centering
\includegraphics[width = 0.95\linewidth]{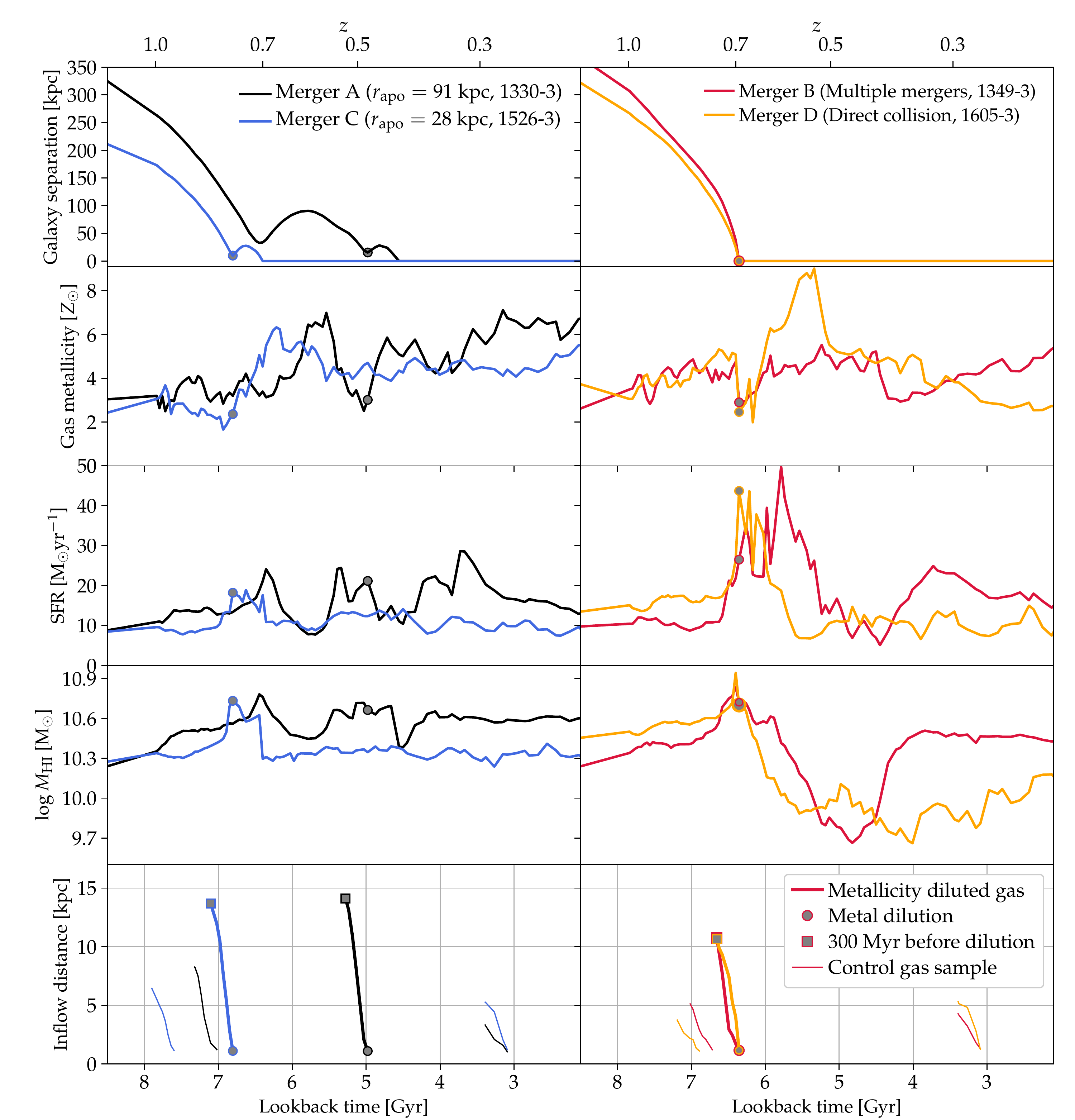}
\caption{At the pericentric passages and at coalescence, we search for an epoch with a diluted central gas metallicity. By comparing the galaxy separation (in the upper panel) with the SFR-averaged gas metallicity inside 1.5 kpc, we identify the pericentric passage or coalescence with the strongest metal dilution. The corresponding time is marked with a grey-filled circle. We also show the SFR and the \iona{H}{i} mass. In the lower panel, we trace gas cells in the metal-diluted gas reservoir back in time to quantify the inflows. The thick-solid lines show the median distance to the galaxy centres in the last 300 Myr leading up to the metallicity dilution epoch. We compare with \emph{control samples} of gas selected at an earlier or later time (see the thin-solid lines). The gas in the centre of a metallicity-diluted merging galaxy has experienced much more rapid inflows in comparison to normal star-forming galaxies (as quantified by our control gas samples).}
\label{MS_MetallicityDilution}
\end{figure*}

In idealised merger simulations the transfer of gas from the CGM to the ISM (and accretion of gas into galaxy haloes) is not accounted for, but our analysis shows that the CGM strongly contributes to star formation during the merger and in the merger remnant. Therefore, idealised simulations have most likely over-estimated the role of mergers in quenching galaxies, because the replenishment of gas from the CGM is not accounted for. We obtained similar conclusions in \citet{2017MNRAS.470.3946S}, but in this section we have quantitatively determined the main reason to be the gas accretion from the CGM and the halo surroundings.

We have tested how sensitive our quantitative results are to the ISM and CGM definitions. For example, with our above definition dense gas in a satellite galaxy residing at $0.2 R_{200}<r\leq R_{200}$ would be characterised as being part of the CGM, but it could be argued that an ISM designation would also be appropriate. We have repeated the analysis presented in Fig.~\ref{MS_PlotTracedSFR} by including all dense gas with $n>0.13$ cm$^{-3}$ at $0.2 R_{200}<r\leq R_{200}$ to the accounted ISM gas reservoir. For simulation A, C and D this only changes the contributions of ISM and CGM by a few percentage points. For simulation B multiple galaxies are participating in the merger, so here the sensitivity to the CGM definition is more important. With this new definition, the ISM fraction is increased and the CGM fraction is decreased in and after the merger. However, for typical mergers with only two galaxies participating our quantities are only mildly sensitive to the CGM definition.

\subsection{Dilution of gas metallicity}

Observations reveal that galaxy pairs have a lower metallicity compared to field galaxies \citep{2012MNRAS.426..549S} and that this lower metallicity is sustained even after coalescence \citep{2008MNRAS.386L..82M,2010ApJ...723.1255R,2013MNRAS.435.3627E}. Simulations also predict the existence of metallicity dilution in mergers \citep{2010ApJ...710L.156R}. Furthermore, integral field spectroscopy has recently established the spatially resolved nature of the gas metallicity dilution \citep{2019MNRAS.482L..55T}. Observations also find the metallicity dilution in galaxy mergers \citep{10.1093/mnras/stv1232} to be stronger than predicted by the \emph{fundamental metallicity relation} \citep[FMR;][]{2008ApJ...672L.107E,2010A&A...521L..53L,2010MNRAS.408.2115M}, which is a two-dimensional connection between SFR, gas metallicity, and the stellar mass of galaxies. The reason is suggested to be faster and more violent inflows \citep{2020MNRAS.494.3469B} in comparison to normal star-forming galaxies. We now proceed with a quantitative analysis of the gas flows in our simulations to test this scenario.

In Fig.~\ref{MS_MetallicityDilution} we study the metallicity dilution, and how it relates to other galaxy quantities. We show the separation of the galaxies (same as shown in Fig.~\ref{MS_Orbits}), the SFR-averaged gas metallicity measured in the central 1.5 kpc, the SFR (same as in Fig.~\ref{MS_Orbits}), and the \iona{H}{i} mass within $0.2R_{200}$. To identify epochs of metallicity dilution, we search for a dip in metallicity at the pericentric passages and final coalescence, where we expect the metallicity dilution to be strongest. The time of the strongest dilution is marked with a grey circle. As seen in the figure, the central gas metallicity is a fluctuating quantity, which is non-trivially linked to the orbit. There is an anti-correlation between gas metallicity and \iona{H}{i} mass, since inflows add low-metallicity gas to the galaxy. Indeed, our identified metallicity dilution epoch occurs in all simulations close to the time of a local maximum for the \iona{H}{i} gas mass. At the same time, a high SFR is also connected to an increase in the \iona{H}{i} -- this is expected, because our ISM model \citep{2003MNRAS.339..289S} is tuned to re-produce the observed correlation between gas surface density and SFR surface density \citep{1989ApJ...344..685K}. On a similar note, our analysis agrees with \citet{2018MNRAS.477L..16T}, who identified a comparable variability time-scale for the SFR and the gas metallicity.

To quantitatively study properties of inflows we select the metallicity diluted gas at the time, where we identify the strongest dilution, and track the flow back in time using gas tracer particles. For each gas tracer, we determine its distance to the merger remnant's progenitor galaxies ($r_1$ and $r_2$). In the last 300 Myr before the metallicity dilution epoch we show the evolution of the median value of $\min(r_1,r_2)$ -- see the lower panel of Fig.~\ref{MS_MetallicityDilution}. 300 Myr before the dilution epoch the median distance of the gas cells is 10--15 kpc for the four galaxies that we have simulated.

To generate a \emph{control sample} indicating the behaviour of this quantity for normal star-forming galaxies, we plot the same quantity well before the first pericentric passage and well after the final coalescence (see thin solid lines in the lower panel of Fig.~\ref{MS_MetallicityDilution}). With these choices, the controls are representative of normal star-forming galaxies not undergoing a merger. For the control samples, we do not see median distances above 8 kpc, with the most common value being around 5 kpc.

Based on the determined \emph{inflow distance}, we can construct a time-scale for gas entering the centre from a fixed distance. We can, for example, construct a time-scale for gas to migrate from 10 kpc to the central 1.5 kpc. This time-scale is 200--300 Myr for the metallicity dilution epochs, and 450--1000 Myr for the control samples representing normal star-forming galaxies. We conclude that metallicity-diluted mergers experience more rapid inflows in comparison to the normal galaxy population, and our analysis hence supports the scenario from \citet{2020MNRAS.494.3469B}. We caution that our claims that the metallicity dilution is associated with rapid inflows only apply to low redshift, as our analysis has been performed at $z\lesssim 0.7$. It also remains undetermined from an observational point of view as to whether the gas metallicity of mergers at higher redshift is captured by the FMR, as shown by \citet{2021MNRAS.501..137H}.

Simple analytical galaxy models have successfully modelled the quantities entering the FMR \citep{2012MNRAS.421...98D,2013ApJ...772..119L}. We do, however, note that such models only capture processes in normal star-forming galaxies and not merger events with rapid inflows, which may generally be outliers to the relation.

\section{Discussion: Why are the merger remnants \new{still} star-forming?}\label{Discussion}

Major mergers are often \new{expected to quench} galaxies, yet the merger remnants in our simulations are all star-forming. \new{This raises the question: why?} All the mergers studied here are, however, wet (gas-rich), so this makes the lack of quenching less surprising. Also, we have demonstrated that a key for the galaxies to maintain a high star formation rate after the merger is gas, which resided in the CGM ahead of the merger. Most idealised simulations have not included this gas reservoir, implying that they have probably overestimated the importance of quenching in mergers. Another possibility is that a fraction of previous generations of idealised simulations used more efficient AGN feedback. Our AGN feedback model, like most feedback models used in cosmological simulations, is calibrated to reproduce observed relations originating from abundance matching or observed scalings. With such a calibration we do not achieve quenching in mergers.

\new{Our galaxy formation model is able to produce (Milky-Way-mass-)galaxies being quenched or having an elliptical stellar structure \citep{2017MNRAS.467..179G}. It strengthens our conclusion that a merger may not imply an elliptical remnant. Our model has, however, not been applied to cosmological box simulations, so it is unclear whether the model is over- or under-producing the fraction of quenched ellipticals in different environments.}

Other cosmological simulations, which found merger remnants with similar mass haloes ($M_{200}\simeq 10^{12}$ M$_\odot$) to be star-forming, include the analysis of the disc of merger remnants in the Illustris simulation \citep{2017MNRAS.467.3083R}, the study of post-mergers in Illustris-TNG \citep{2020MNRAS.493.3716H,2021MNRAS.504.1888Q}, the mergers in the Auriga simulations \citep{2018MNRAS.479.3381B}, and the VINTERGATAN simulation \citep{2020arXiv200606008A,2020arXiv200606011R}. There is therefore a consensus in simulations that merger remnants (at low redshift, having a Milky-Way-like mass) \new{remain star-forming}.

Observations also identify star-forming merger remnants. \citet{2016ApJ...830..137A}, for example, found a reservoir of molecular gas and a normal star formation rate $0.5$ Gyr after a merger, and  \citet{2018MNRAS.478.3447E} found notable \iona{H}{i} reservoirs in post-mergers, implying that the \emph{gas-blowout-phase} from the scenario of \citet{2008ApJS..175..356H} is questionable. We note that a significant amount of \iona{H}{i} gas both during and after a merger is consistent with our simulations (see Fig.~\ref{MS_MetallicityDilution}).

Furthermore, the Galaxy Zoo project found mergers not to be the main pathway to quenching \citep{2017ApJ...845..145W}. While merging galaxies approach each other, their molecular gas mass is also consistent with normal galaxies \citep{2018MNRAS.476.2591V,2018ApJ...868..132P}, further supporting a scenario, where mergers are not evacuating gas from galaxies.

Even though mergers are not necessarily the main pathway to quenching, it is possible that mergers may be efficient in quenching galaxies at higher masses or at higher redshift. \citet{2019ApJ...874...17B}, for example, observed quenched galaxies at $z\gtrsim 1.5$ and identified a slow and a rapid mode of quenching, where only the latter is expected to be associated to mergers (or a starburst event).

\section{Conclusion}\label{Conclusion}

We have used cosmological simulations to study gas flows in and around gas-rich major mergers. Galaxy formation models predict continuous transfer of gas between the ISM and CGM of galaxies, and we have shown here that mergers are no exception to this phenomenon. Our main conclusions are:

\begin{itemize}
\item At the time of coalescence of the merging galaxies, there is a contribution of 27-51 per cent to the SFR from gas accreted from the CGM during the merger. Idealised merger simulations without a CGM therefore underestimate the SFR of the starburst associated with the coalescence. The gas accreted from the CGM and the cosmological surroundings is also important for maintaining and reigniting star formation in merger remnants. At $z=0$ only 24-45 per cent of the SFR comes from gas which was in the ISM prior to the merger.
\item The gas in bridge primarily comes from the ISM of the two progenitor galaxies (47--67 of the gas was in the ISM 1 Gyr ahead of being in the disc), but our simulations also reveal a significant contribution (33--48 per cent) coming from the CGM.
\item The gas in bridges between mergers is supersonic turbulent with a Mach number of \new{1.6--3.3} at the midpoint in between the galaxies. This is important, because star formation theories predict supersonic turbulent clouds to have a lower star formation efficiency in comparison to gas in galaxy discs. 

\item We identify gas metallicity dilution in the mergers. We confirm the suggestion from \citet{2020MNRAS.494.3469B} that the central metallicity diluted gas has experienced rapid inflows in comparison to normal star-forming galaxies. This is the reason why the mergers are not captured by the fundamental metallicity relation (FMR). All our mergers are at a relatively low redshift ($z\lesssim 0.7$), so this result only applies to low redshift. Observationally, it has also not been established whether or not the FMR well captures the behaviour of high redshift mergers. Understanding major mergers in the context of the FMR at high redshift remains a challenge for future work.
\item In merger events, the \iona{H}{i}-averaged line-of-sight velocity goes from revealing signatures characteristic for a rotating disc to showing a complex clumpy structure. Just as expected for the orbit of stars, gas is randomised in the mergers.
\end{itemize}

\new{The results in this paper rely on galaxies merging between $z=0.3$ and $z=0.8$. Our quantitative results will therefore not necessarily hold for galaxies outside this redshift range. A further inspection of the progenitors reveals that the merging galaxies have a disc morphology, e.g. with  visible spiral arms. We therefore expect our results to apply to \emph{low-redshift, gas-rich galaxies}. At higher redshift dense clumps \citep{2009MNRAS.397L..64A,2010MNRAS.404.2151C,2014ApJ...780...57B,2017MNRAS.468.3628B} and bursty star formation \citep{2017MNRAS.466...88S,2018MNRAS.473.3717F,2021MNRAS.501.4812F} are important, so it is unclear whether gas accretion in mergers occurs on similar time-scale as at lower redshift. Further simulations would therefore be necessary to address mergers at higher redshift.}

\section*{Acknowledgements}

\new{We thank the referee for insightful comments}. This paper is a result of interesting discussions initiated at the workshop, \emph{Galaxy interactions and mergers across cosmic time} in Sexten, Italy, March 2018. MS and CP acknowledges support by the European Research Council under ERC-CoG grant CRAGSMAN-646955. MV acknowledges support through NASA ATP grants 16-ATP16-0167, 19-ATP19-0019, 19-ATP19-0020, 19-ATP19-0167, and NSF grants AST-1814053, AST-1814259,  AST-1909831 and AST-2007355. MHH acknowledges support from the William and Caroline Herschel Postdoctoral Fellowship Fund. We thank the contributors and developers to the software packages \textsc{yt} \citep{2011ApJS..192....9T} and \textsc{astropy} \citep{2018AJ....156..123A}, which we have used for the analysis in this paper.

\section*{Data availability}
The data and scripts for this article will be shared on reasonable request to the corresponding author. The \textsc{arepo} code is publicly available.

\footnotesize{
\bibliographystyle{mnras}
\bibliography{ref}
}

\appendix
%

\section{Stellar maps}\label{StellarMaps}

\begin{figure*}
\centering
\includegraphics[width = 1.0\linewidth]{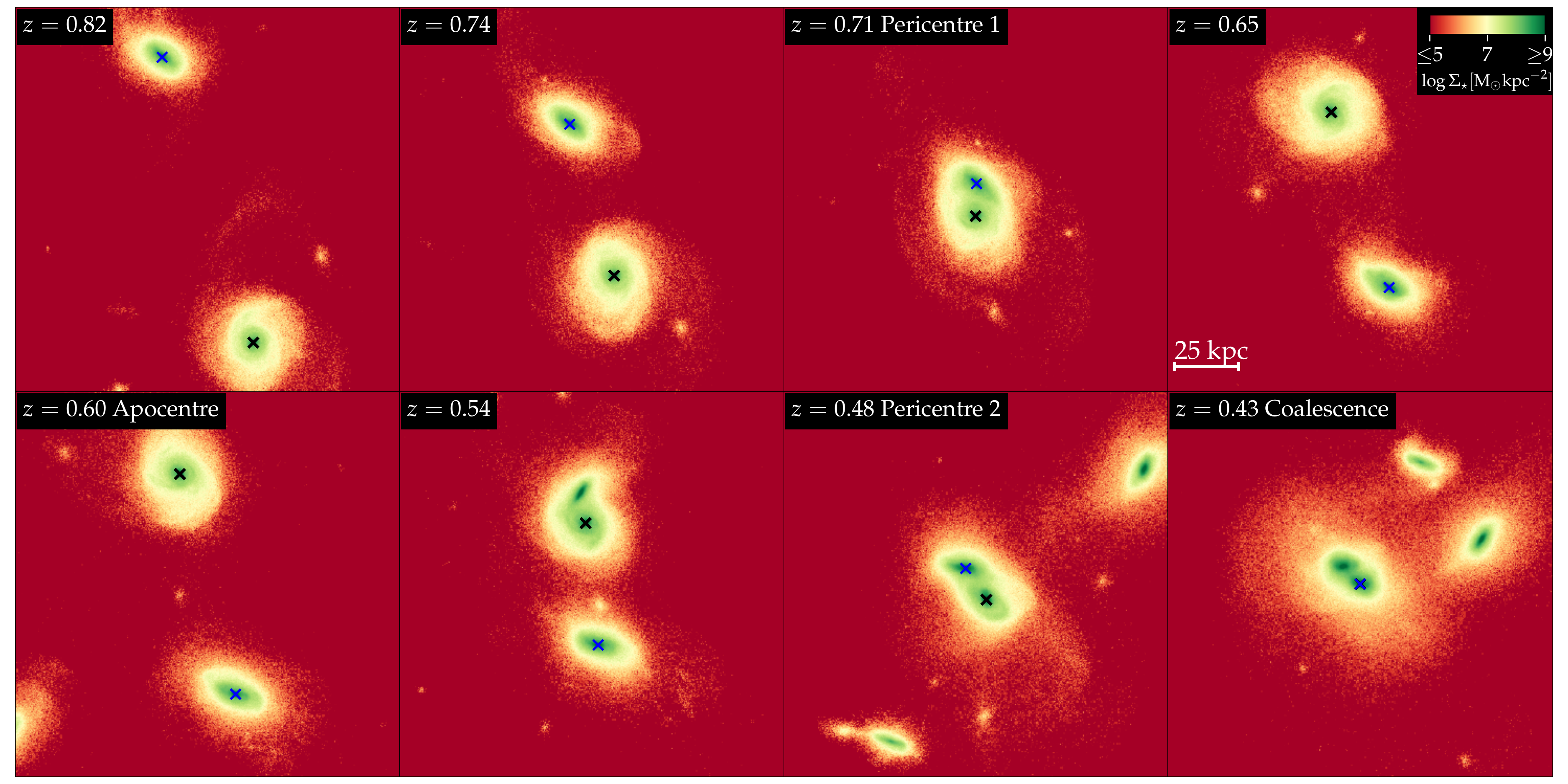}
\caption{The stellar surface density of merger A. The \iona{H}{i} bridge at the apocentre (see Fig.~\ref{Fig100_HI_sequence_1330-3MHD}) is not accompanied by a visual stellar bridge. This shows that the collisional nature of the gas behaves significantly different than the collisionless nature of the stars.}
\label{Fig104_Stars_sequence_1330-3MHD}
\end{figure*}

\begin{figure*}
\centering
\includegraphics[width = 1.0\linewidth]{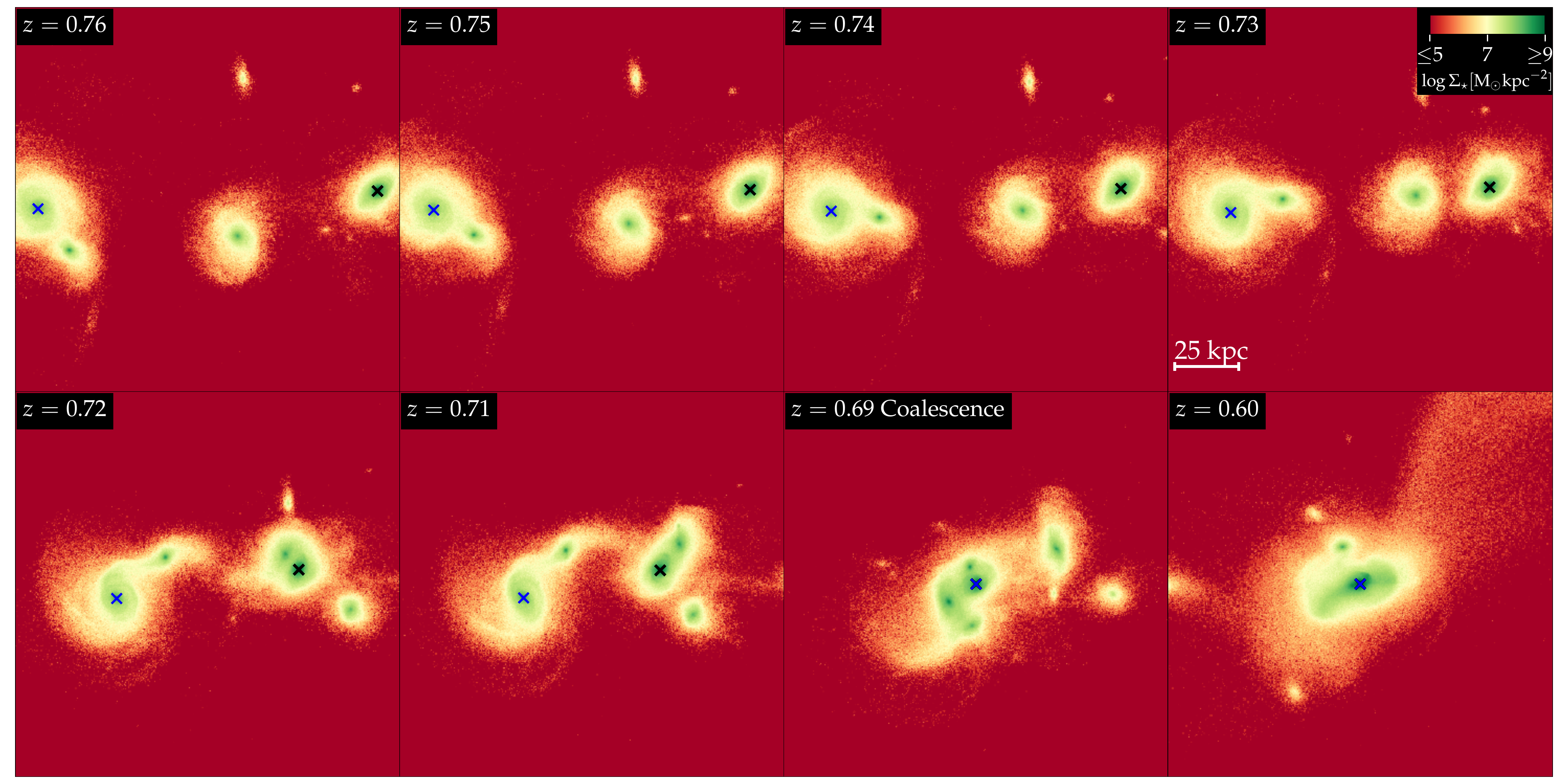}
\caption{Same as Fig.~\ref{Fig104_Stars_sequence_1330-3MHD} but for merger B. At $z=0.71$ a gaseous structure connecting the galaxies is visible in \iona{H}{i} (Fig.~\ref{Fig100_HI_sequence_1349-3MHD}), but this is not visible for the stellar distribution. The stripped distribution of stars in the upper right part of the $z=0.60$ panel is also visible in \iona{H}{i}, implying that tidal forces are responsible for this stripping.}
\label{Fig104_Stars_sequence_1349-3MHD}
\end{figure*}

\begin{figure*}
\centering
\includegraphics[width = 1.0\linewidth]{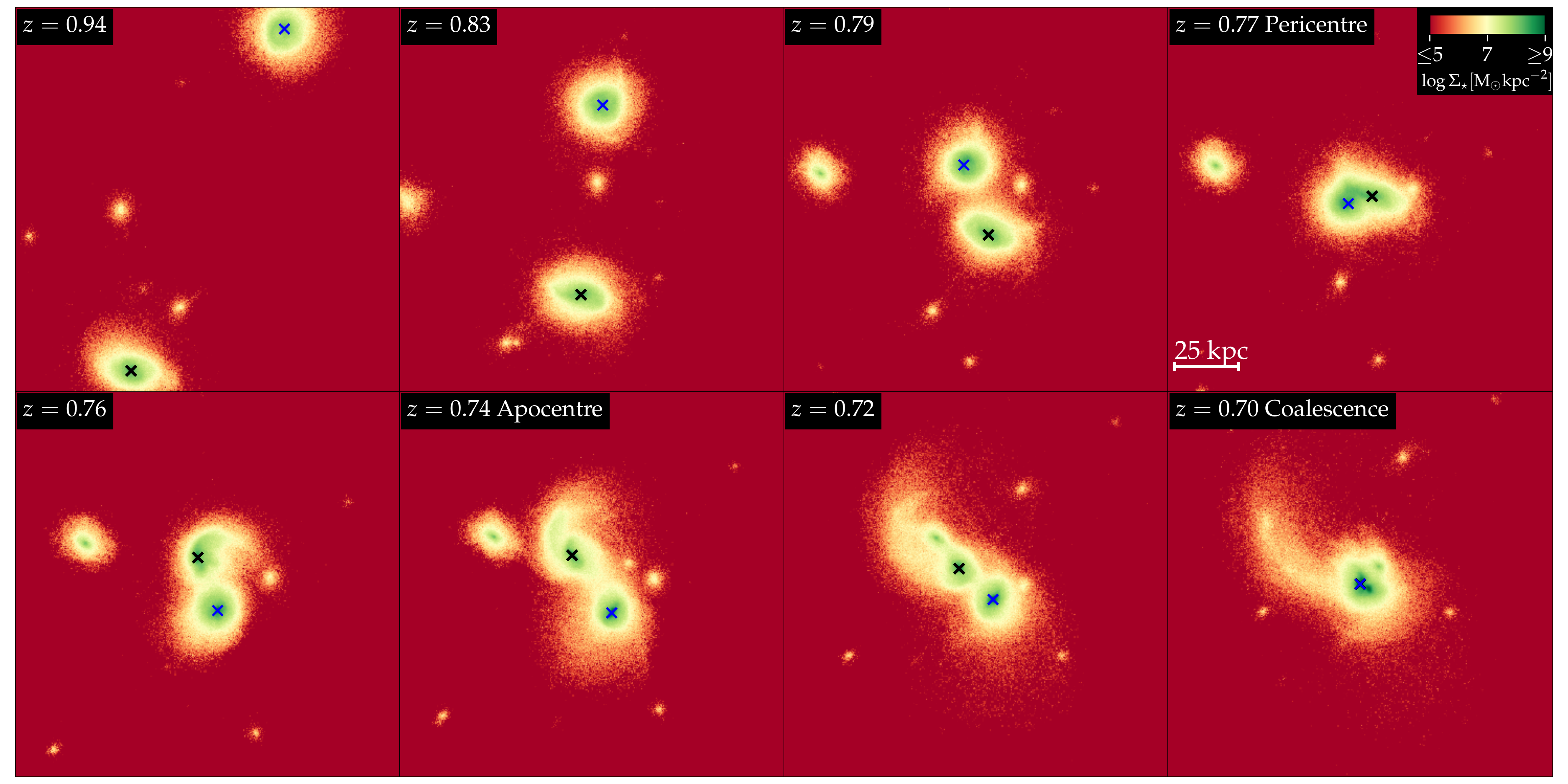}
\caption{Same as Fig.~\ref{Fig104_Stars_sequence_1330-3MHD} but for merger C. The \iona{H}{i} bridge seen prior to the pericentric passage at $z=0.79$ (Fig.~\ref{Fig100_HI_sequence_1526-3MHD}) is not visible in the stellar distribution, again implying that collisional processes are responsible for this features. At the time of the apocentre there is both a bridge seen in \iona{H}{i} and the stellar distribution.}
\label{Fig104_Stars_sequence_1526-3MHD}
\end{figure*}

\begin{figure*}
\centering
\includegraphics[width = 1.0\linewidth]{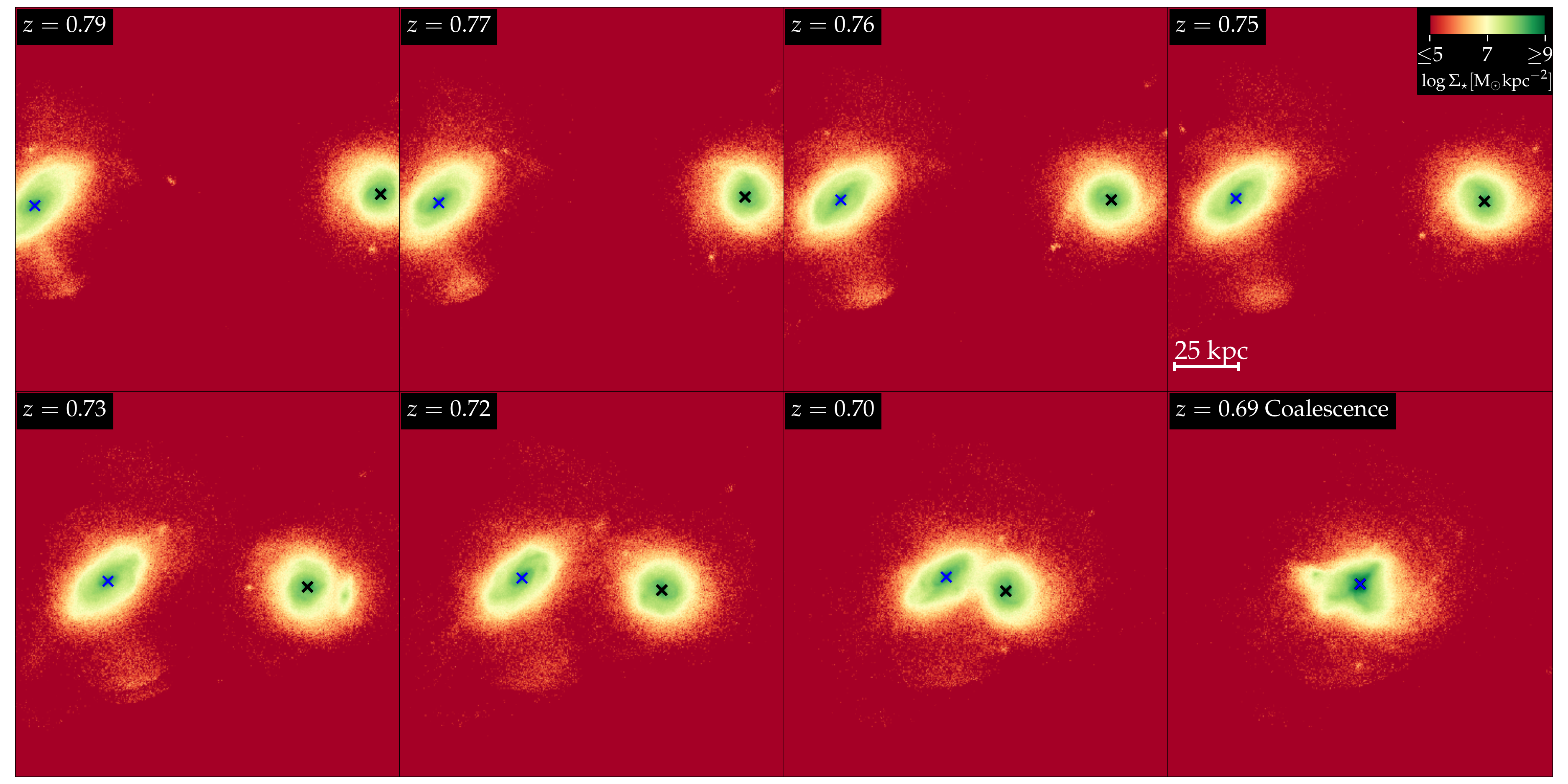}
\caption{Same as Fig.~\ref{Fig104_Stars_sequence_1330-3MHD} but for merger D. At $z=0.73$ and $z=0.72$ the \iona{H}{i} distribution reveals dense gas between the galaxies, but this is not accompanied by a similar increase in stellar surface density. The reason is again collisional processes, or alternatively it could be caused by gas accretion, as studied in Fig.~\ref{Fig100_HI_sequence_TracedGasInBridge}.}
\label{Fig104_Stars_sequence_1605-3MHD}
\end{figure*}

\new{Fig.~\ref{Fig104_Stars_sequence_1330-3MHD}-\ref{Fig104_Stars_sequence_1605-3MHD} show stellar surface density maps during the merger for our four simulations. By comparing these to our \iona{H}{i} maps (Fig.~\ref{Fig100_HI_sequence_1330-3MHD}-\ref{Fig100_HI_sequence_1605-3MHD}), we can assess the origin of the bridges connecting the merging galaxies.} 

\new{For merger A, there is a visually remarkable bridge in the \iona{H}{i} map at the time of the apocentre, but in the stellar map this is absent (Fig.~\ref{Fig104_Stars_sequence_1330-3MHD}). This shows that collisional processes (e.g., shocks and ram pressure) and gas accretion may be important for the formation of a gas bridge, since a stellar bridge would be prominent if tidal stripping from the disc was the main mechanism. For merger B we reach the same conclusion for the gas between the two main progenitors at $z=0.71$ (Fig.~\ref{Fig104_Stars_sequence_1349-3MHD}).}

\new{At the time of the apocentre in merger C, there is a visual bridge present in both the \iona{H}{i} column density and stellar surface density (Fig.~\ref{Fig104_Stars_sequence_1526-3MHD}). Tidal stripping of gas and stars from the disc therefore plays an important role (as in the bridge formation scenario presented in \citealt{1972ApJ...178..623T}). Prior to the pericentric passage at $z=0.79$ the \iona{H}{i} gas reveals a bridge between the galaxies, but this is absent in the stellar distribution, so here we expect collisional processes to be important. For merger D, the same is the case at $z=0.72$ ahead of the coalescence (Fig.~\ref{Fig104_Stars_sequence_1605-3MHD}).}

\bsp	
\label{lastpage}
\end{document}